\documentclass[aps,prc,onecolumn,showpacs,showkeys,preprintnumbers,amsmath,amssymb,a4paper]{revtex4-1}
\usepackage{amssymb}
\usepackage{graphics}
\usepackage{psfrag}
\usepackage{epsfig}
\usepackage{multirow,mathtools}
\usepackage{bm}
\def\bea{\begin{eqnarray}}
\def\eea{\end{eqnarray}}
\def\beq{\begin{equation}}
\def\eeq{\end{equation}}
\def\nn{\nonumber}
\def\bp{\mathbf{p}}

\def\bq{\mathbf{q}}
\def\br{\mathbf{r}}
\def\bL{\mathbf{L}}
\def\bS{\mathbf{S}}

\def\bsigma{\bm{\sigma}}
\def\brho{\bm{\rho}}
\def\blam{\bm{\lambda}}
\newcommand{\pbarp}{{\bar p p}}
\newcommand{\nbarn}{{\bar N  N}}
\newcommand{\ybary}{{\bar YY}}
\newcommand{\ycbaryc}{{\bar Y_cY_c}}
\newcommand{\lbarl}{{\bar \Lambda \Lambda}}
\newcommand{\lcbarlc}{\bar{\Lambda}_c^- {\Lambda}_c^+}
\newcommand{\lbars}{{\bar \Lambda \Sigma}}

\newcommand{\lcbarsc}{\bar{\Lambda}_c^- {\Sigma}_c^+}
\newcommand{\scbarlc}{\bar{\Sigma}_c^- {\Lambda}_c^+}
\newcommand{\sbars}{{\bar \Sigma \Sigma}}
\newcommand{\spbarsp}{{\bar \Sigma^- \Sigma^+}}
\newcommand{\sobarso}{{\bar \Sigma^0 \Sigma^0}}
\newcommand{\smbarsm}{{\bar \Sigma^+ \Sigma^-}}
\newcommand{\scbarsc}{{\bar \Sigma_c \Sigma_c}}
\newcommand{\scpbarscp}{{\bar \Sigma_c^{--} \Sigma_c^{++}}}
\newcommand{\scobarsco}{{\bar \Sigma_c^{-} \Sigma_c^{+}}}
\newcommand{\scmbarscm}{{\bar \Sigma_c^{0} \Sigma_c^{0}}}
\newcommand{\xbarx}{{\bar \Xi \Xi}}
\newcommand{\xobarxo}{{\bar \Xi^0 \Xi^0}}
\newcommand{\xmbarxm}{{\bar \Xi^+ \Xi^-}}
\newcommand{\xcbarxc}{{\bar \Xi_c \Xi_c}}
\newcommand{\xcobarxco}{{\bar \Xi_c^0 \Xi_c^0}}
\newcommand{\xcpbarxcp}{{\bar \Xi_c^- \Xi_c^+}}
\newcommand{\xcobarxcop}{{\bar \Xi_c^{\prime 0} \Xi_c^{\prime 0}}}
\newcommand{\xcpbarxcpp}{{\bar \Xi_c^{\prime -} \Xi_c^{\prime +}}}
\begin{document}
\title{Production of charmed baryons in $\bar p p$ collisions close to their thresholds}
\author{J. Haidenbauer$^1$ and G. Krein$^2$}
\affiliation{
$^1$Institute for Advanced Simulation, Institut f\"ur Kernphysik, and J\"ulich Center for Hadron
Physics, Forschungszentrum J\"ulich, D-52425 J\"ulich, Germany \\
$^2$Instituto de F\'{\i}sica Te\'{o}rica, Universidade Estadual
Paulista,
Rua Dr. Bento Teobaldo Ferraz, 271 - Bloco II, 01140-070 S\~{a}o Paulo, SP, Brazil
}

\begin{abstract}
Cross sections for the charm-production reactions $\bar p p \to \bar \Lambda_c^- \Sigma_c^+$, 
$\bar \Sigma_c\Sigma_c$, $\bar \Xi_c\Xi_c$, and $\bar \Xi_c'\Xi_c'$ 
are presented, for energies near their 
respective thresholds. The results are based on a calculation performed 
in the meson-exchange framework in close analogy to earlier studies of
the J\"ulich group on the strangeness-production reactions 
$\bar p p \to \bar \Lambda\Sigma$, $\bar \Sigma\Sigma$, $\bar \Xi\Xi$ 
by connecting the two sectors via SU(4) flavor symmetry. 
The cross sections are found to be in the order of $0.1 - 1$ $\mu b$  
at energies of $100$ MeV above the respective thresholds, for all 
considered channels. 
Complementary to meson-exchange, where the charmed baryons are
produced by the exchange of $D$ and $D^*$ mesons, a charm-production 
potential derived in a quark model is employed for assessing uncertainties.
The cross sections predicted within that picture turned out to be significantly
smaller. 
\end{abstract}


\maketitle

\section{Introduction}
\label{sec:intro}

The FAIR project at the GSI laboratory has an 
extensive program aiming at a high-accuracy spectroscopy of charmed hadrons 
and at an investigation of their interactions with ordinary matter \cite{PANDA}. 
For the feasibility of such studies, specifically those of the {$\rm \bar P$}ANDA experiment
\cite{Wiedner:2011} 
the production rate for charmed hadrons in antiproton-proton ($\pbarp$) is a key factor. 
Knowledge of such production rate is also relevant for studies of in-medium changes 
of charmed hadrons{\textemdash}for recent references, see 
Refs.~\cite{in-medium1,in-medium2,Shyam:2016}. 
However, presently very little is known about the strength of the interaction of charmed
hadrons with ordinary baryons and mesons. In view of that,  
over the last few years we have looked at the exclusive charm-production reactions 
$\pbarp \rightarrow \lcbarlc$ \cite{Haidenbauer:2010} and $\pbarp \rightarrow D\bar D,\,
D_s\bar D_s$ \cite{Haidenbauer:2014,Haidenbauer:2015}  
close to their thresholds with the objective to provide with our predictions 
estimations for the pertinent cross sections. 
 
In the present paper we extend our study of the reaction $\pbarp \rightarrow \lcbarlc$
\cite{Haidenbauer:2010} to the production of other charmed baryons such as 
the $\Sigma_c$, the $\Xi_c$ and the $\Xi_c'$. 
The projected antiproton beam-momentum available for the {$\rm \bar P$}ANDA experiment
reaches up to $15$ GeV/c corresponding to a center-of-mass energy of $\sqrt{s} = 5.5$ GeV \cite{Boca}. 
Thus, the production of most of the charmed members of the lowest SU(4) 
$J^P = 1/2^+$ baryon $20$-plet is possible at {$\rm \bar P$}ANDA, including 
the $\Xi_c$ and $\Xi_c'$ and even the $\Omega_c^0$ \cite{PDG}. 
While there is a large number of calculations for $\pbarp \rightarrow \lcbarlc$ 
\cite{Kroll:1988cd,Kaidalov:1994mda,Titov:2008,Goritschnig:2009sq,Khodjamirian:2012,Shyam:2014,Wang:2016}
this can not be said for the production of other charmed baryons.
Khodjamirian et al. \cite{Khodjamirian:2012} published cross 
sections for $\lcbarsc$ and $\scbarsc$. 
Titov and K\"ampfer provided results for $d\sigma / dt$, for $\lcbarsc$ and $\scbarsc$ 
in \cite{Titov:2008} and for $\xcbarxc$ in \cite{Titov:2011}. 
% In the latter work two step processes (loop diagrams) with intermediate
% $\lbarl$ and $\lcbarlc$ states are considered. 
However, their analysis focusses on the region of small momentum transfer and integrated 
cross sections are not given. 
The earliest study we are aware of where integrated cross sections for $ \scpbarscp$ were
presented is the one by Kroll et al.~\cite{Kroll:1988cd}.
%
% Genz et al.~\cite{Genz:1991} study $\bar p p$ reactions via double annihilation 
% of quarks, also annihilation into $\bar\Xi\Xi$.

Our analysis of charm production is done in complete analogy to that 
of the reactions $\pbarp \rightarrow \lbarl,\lbars,\sbars,\xbarx$
performed by the J\"ulich group some time ago 
\cite{Haidenbauer:1991,Haidenbauer:1992,Haidenbauer:1993,Haidenbauer:XX}.
In those studies the hyperon-production 
reaction is considered within a coupled-channel model. This allows to take into
account rigorously the effects of the initial ($\pbarp$) and final ($\ybary$)
state interactions which play an important role for energies near the
production threshold 
\cite{Haidenbauer:1991,Haidenbauer:1992,Kohno,Alberg}. 
The microscopic strangeness production is described by meson exchange 
and arises from the exchange of the $K$ and $K^*$ mesons. 
The elastic parts of the interactions in the initial ($\pbarp$) and final ($\ybary$) states 
are likewise described by meson exchange, while annihilation processes 
are accounted for by phenomenological optical potentials. Specifically,
the elastic parts of the initial- (ISI) and final-state interactions (FSI)
are $G$-parity transforms of an one-boson-exchange variant of the
Bonn $NN$ potential~\cite{obepf} and of the hyperon-nucleon model~A of 
Ref.~\cite{Holzenkamp89}, respectively. 
With this model a good overall description of the 
$\pbarp \to \lbarl$, $\pbarp \to \lbars {\rm + c.c.}$, and $\pbarp \to \sbars$ 
data obtained in the P185 experiment at LEAR (CERN) 
\cite{PS185} could be achieved and its results are also in line with the scarce 
experimental information for $\pbarp \to \xbarx$ \cite{Haidenbauer:XX}. 

The extension of the model to the charm sector is based on SU(4) flavor symmetry. 
Accordingly, the elementary charm
production process is described by $t$-channel $D$ and $D^*$ meson exchanges. 
Note that the symmetry is invoked primarily as guideline for providing constraints 
on the pertinent baryon-meson coupling constants. 
Though we do not expect that the SU(4) symmetry should hold, recent calculations 
of the relevant coupling constants within QCD light-cone sum rules suggest that 
the actual deviation from the SU(4) values could be only in the order of a 
factor $2$ or even less \cite{Khodjamirian:2012}; even smaller deviations have
been obtained~\cite{3p0} in a constituent quark model calculation using the 
$^3P_0$ pair-creation mechanism.
We examine the sensitivity of the results to variations in the elastic and 
annihilation parts of the initial $\pbarp$ interaction. Furthermore, 
as already done for $\lcbarlc$ \cite{Haidenbauer:2010}, we investigate the effect 
of replacing the meson-exchange transition by a charm-production potential derived 
in a quark model. Again this serves for assessing uncertainties in the model, since 
one could question the validity of a meson-exchange description of the transition 
in view of the large masses of the exchanged mesons.  
In this context we want to note that meson-exchange as well as the quark model lead 
to rather short ranged transition potentials. Thus, practically speaking those can be
viewed as being contact interactions where the pertinent coupling constants
are simply saturated \cite{Epelbaum:2001} by the dynamics underlying the two considered 
approaches.

In the next two Sections we introduce the basic ingredients of the model. 
In Section~\ref{sec:res} we present numerical results for total cross sections 
for the various $\ycbaryc$ channels, utilizing for the charm-production
mechanism meson-exchange as well as the quark model.  
A summary of our work is presented in Section~\ref{sec:sum}.
Details on the transition potential in the quark model and on the
SU(4) coupling constants that enter the meson-exchange transition potential
are collected in Appendices. 

%%%%%%%%%%%%%%%%%%%%%%%%%%%%%%%%%%%%%%%%%%%%%%%%%%%%%%%
\section{The model}
\label{sec:model}

We calculate the charm production reactions $\pbarp \rightarrow \bar Y_c Y_c$
in close analogy to the original J\"ulich coupled channel 
approach~\cite{Haidenbauer:1991,Haidenbauer:1992,Haidenbauer:1993,Haidenbauer:XX} 
to strangeness production. The transition amplitude is obtained from the 
solution of a multi-channel Lippmann-Schwinger (LS) equation,
\begin{eqnarray}
\boldsymbol{T} &=& \boldsymbol{V} + \boldsymbol{V \, G_0 \, T},
\label{TmatM}
\end{eqnarray}
which reads explicitly in terms of the channels $\mu$ ($\nu$) corresponding to
$\nbarn$, $\lcbarlc$, $\lcbarsc$, $\scbarlc$, $\scbarsc$, and $\xcbarxc$,
\begin{eqnarray}
T^{\mu\nu}(\bp_\mu,\bp_\nu,z) &=& V^{\mu\nu}(\bp_\mu,\bp_\nu,z) + \sum_{\gamma}
\int d^3p_\gamma \, V^{\mu\gamma}(\bp_\mu,\bp_\gamma,z) \, 
G_{0}^{\gamma}(\bp_\gamma,z) \, T^{\gamma\nu}(\bp_\gamma,\bp_\nu,z) ,
\label{Tmat}
\end{eqnarray}
Here $z$ is the total energy and $\bp_\nu$ ($\bp_\mu$) the relative 
momentum in the initial (final) state in the center-of-mass.
The propagator, $G_0(\bp,z)$, is given by 
\begin{equation}
G_0^{\gamma}(\bp_\gamma,z) = 1/(z - E^{\gamma}_{\bp_\gamma} + i\epsilon) 
\label{G0}
\end{equation}
with $E^{1}_{\bp_1} = E^{\bar N}_{\bp_{\nbarn}} + E^{N}_{\bp_{\nbarn}}$, etc., 
being the relativistic energies of the two baryons in the intermediate state. 
The calculations are performed in isospin basis, which is sufficient
for an exploratory study. Moreover, the mass splitting between 
$\Sigma^{++}_c$, $\Sigma^{+}_c$, and $\Sigma^{0}_c$ is rather small
\cite{PDG}. This is different in the strangeness sector where there
is a sizable mass difference between the 
$\Sigma^{+}$, $\Sigma^{0}$, and $\Sigma^{-}$ which made a calculation
in the particle basis mandatory \cite{Haidenbauer:1993}. 

\begin{figure}[t]
\begin{center}
\includegraphics[height=220mm]{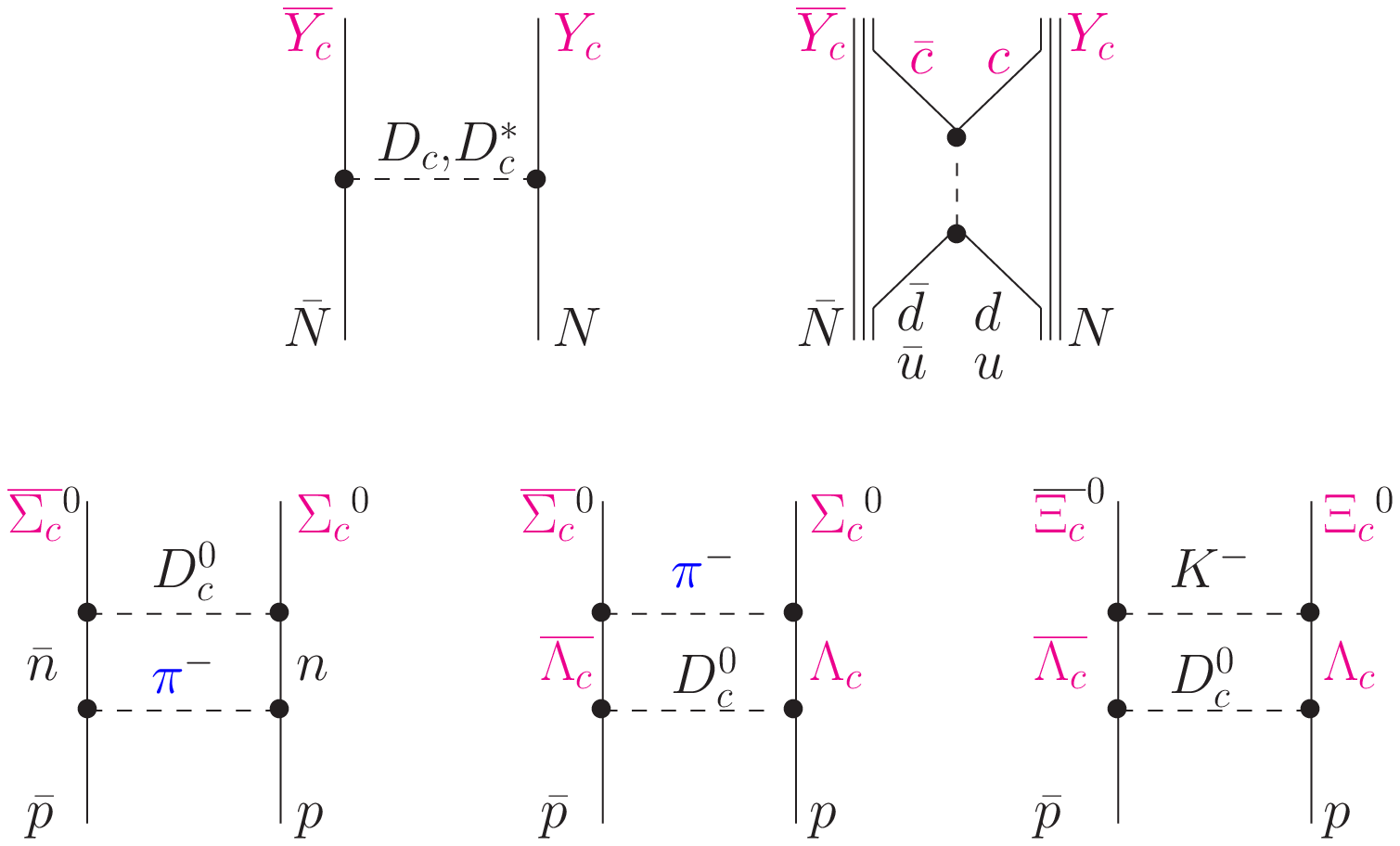}
\vskip -12.5cm
\caption{Upper row: Contributions to the $\bar NN \to \bar Y_c Y_c$ 
transition potential $V^{\mu\nu}$ in the meson-exchange picture (left) and
the quark model (right). 
Lower row: Selected contributions to the $\pbarp \to \bar\Sigma_c^0 \Sigma_c^0$
and $\pbarp \to \bar\Xi_c^0 \Xi_c^0$ transition amplitude generated by the
coupled channel framework. 
}
\label{fig:diag}
\end{center}
\end{figure}

The transition potential from $\nbarn$ to the $\ycbaryc$ channel is given by 
$t$-channel $D$ and $D^*$ exchanges, see Fig.~\ref{fig:diag} (upper row). 
The expressions for the transition potentials are the same as for $K$ and $K^*$ 
exchange and can be found in Ref.~\cite{Holzenkamp89}.
They are of the generic form
\beq
V^{\ycbaryc,\nbarn} (t)
\sim \sum_{M=D,D^*} g_{\bar N\bar Y_c M} g_{NY_c M} 
\frac{F_{\bar N \bar Y_c M}(t) F_{ NY_c M}(t)  }{t-m^2_M} ,
\label{Vtrans}
\eeq
where $g_{NY_c M}$ are coupling constants and $F_{ NY_c M}(t)$
are form factors. Under the assumption of $SU(4)$ symmetry the coupling constants
are identical to those in the corresponding strangeness production reaction 
for $\nbarn \to \lcbarlc$, $\nbarn \to \lcbarsc,\scbarlc$, and $\nbarn \to \scbarsc$,
but differ for $\xcbarxc$, see Appendix \ref{app:su4}. % and Table~\ref{param}. 
With regard to the vertex form factors we use here a monopole form with a cutoff 
mass $\Lambda$ of 3 GeV, at the $NY_cD$ as well as at the $N Y_cD^*$ vertex, as in 
our study of $\lcbarlc$ production~\cite{Haidenbauer:2010}. 

The $\xcbarxc$ channel cannot be reached from $\nbarn$ via single meson
exchange, and the same is also the case for $\bar \Sigma_c^0\Sigma_c^0$
from an initial $\pbarp$ state. The corresponding transition potentials
$V^{\mu\nu}$ are zero. However, the employed coupled-channel framework, 
cf. Eq.~(\ref{Tmat}), generates automatically multistep 
processes so that the corresponding transition amplitudes $T^{\mu\nu}$ are 
no longer zero. Some contributions that arise at the first iteration in the 
LS equation are depicted in the lower row of Fig.~\ref{fig:diag}.  
In principle, there are also contributions from non-iterative two-meson 
exchanges. However, we expect those to be much less important in comparison
to iterated one-meson exchange. In the latter case the two baryons in the 
intermediate states can go on-shell and the pertinent contributions are 
accordingly enhanced \cite{Weinberg}. 

The diagonal potentials $V^{\mu\mu}$ are given by the sum of an elastic part 
and an annihilation part. 
For the $\nbarn$ channel we use again the set of potentials introduced
and described in Refs.~\cite{Haidenbauer:2010,Haidenbauer:2014}. 
Their elastic part is loosely connected (via G-parity transform)
to a simple, energy-independent one-boson-exchange $NN$ potential 
(OBEPF). However, since at the high energies necessary for charm production 
any $NN$ potential has to be considered as being purely phenomenological  
several variants were constructed in order to explore how strongly the results 
on charm production depend on the choice of the $\nbarn$ interaction. 
In two of those variants (called A and A' in \cite{Haidenbauer:2010,Haidenbauer:2014})
only the longest ranged (and model-independent) part of the 
elastic $\nbarn$ interaction, namely one-pion exchange, was kept.  
Model B and C include also some short-range contributions, see the
discussion in \cite{Haidenbauer:2010}. 

\begin{figure}[t]
\begin{center}
\includegraphics[height=75mm,angle=-90]{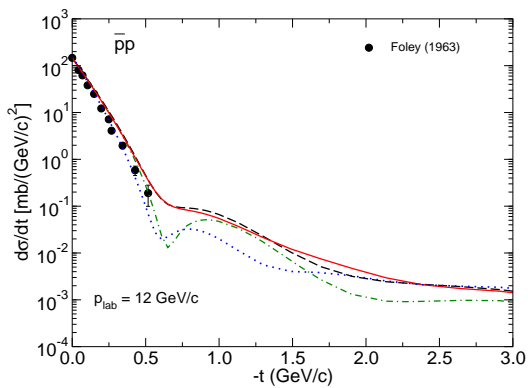}
\caption{Differential cross section for elastic $\pbarp$ scattering at
$p_{lab}$ = 12 GeV/c as a function of $t$. The curves 
represent results based on the $\nbarn$ potentials 
A (dash-dotted line), A' (dotted), B (dashed) and C (solid),
see text for details. 
The experimental information is taken from Foley et al.~\cite{Foley}.
}
\label{pbarp}
\end{center}
\end{figure}

All variants are supplemented by a phenomenological spin-, isospin-, 
and energy-independent optical potential of Gaussian form, in order to take into 
account annihilation,  
\beq
V^{\nbarn \to \nbarn}_{opt}(r)  = (U_0 + i W_0) \, e^{- r^2/2r^2_0} .
\label{Vaa-pp}
\eeq
The free parameters ($U_0$, $W_0$, $r_0$) were determined by a fit to 
$\nbarn$ data in the energy region relevant for the reactions 
$\pbarp \to \lcbarlc$ and $\pbarp \to D\bar D$, i.e. for 
laboratory momenta of $p_{lab} = 6-10$ GeV/c. (Their actual values
can be found in Table~1 of Ref.~\cite{Haidenbauer:2010}.)
The data set comprises total cross sections, and integrated elastic and 
charge-exchange ($\pbarp \to \bar nn$) cross sections.
With all four variants a rather satisfying description of the $\nbarn$ 
data in that energy region could be obtained as documented in 
Refs.~\cite{Haidenbauer:2010,Haidenbauer:2014}. Even at $p_{lab}=12$ GeV/c,
i.e. at a momentum that corresponds roughly to the $\xcbarxc$ threshold, 
the differential cross section is nicely reproduced by all models, 
as exemplified in Fig.~\ref{pbarp}. Evidently, not only the magnitude at 
very forward angles but also the slope is reproduced well by all considered 
$\nbarn$ interactions. 
We want to emphasize that differential cross sections were not included in the 
fitting procedure and are, therefore, predictions of the models. 
 
Note that yet another $\nbarn$ model was considered in \cite{Haidenbauer:2010},
namely Model D, which is based on the full G-parity transformed OBEPF. 
However, its results disagree considerably with the empirical $\pbarp$ differential
cross sections as well as with the integrated charge-exchange cross sections
and, thus, cannot be considered to be realistic. 
Because of that it was no longer utilized in our study of $\pbarp \to D\bar D$ 
\cite{Haidenbauer:2014,Haidenbauer:2015} and we will not use it here either.

In Ref.~\cite{Haidenbauer:2010}
the interaction in the final $\lcbarlc$ system was assumed to be the
same as the one in $\lbarl$. Specifically, this means that the
elastic part of the interaction is fixed by coupling constants and 
vertex form factors taken from the hyperon-nucleon model~A of 
Ref.~\cite{Holzenkamp89}, while the annihilation part is again
parameterized by an optical potential which contains, however, 
spin-orbit and tensor components in addition to a central component \cite{Haidenbauer:1991}:
\bea
V^{\lcbarlc\to\lcbarlc}_{opt}(r)  &=& \bigl[U_c + i W_c + (U_{LS} + i W_{LS}) \, \bL\cdot\bS
\nn \\
&& + \, (U_t + i W_t) \, \bsigma_{\Lambda_c}\cdot\br \, \bsigma_{\bar\Lambda_c}\cdot\br\bigr]
\, e^{- r^2/2r^2_0} .
\label{Vaa-ll}
\eea
The free parameters in the optical $\lbarl$ potential were determined in
Ref.~\cite{Haidenbauer:1991} by a fit to data on total and differential cross sections, and
analyzing power for $\pbarp \to \lbarl$. As already emphasized in \cite{Haidenbauer:2010},
we do not expect that the $\lcbarlc$ interaction will be the same on a quantitative level.
But at least the bulk properties should be similar, because in both cases near threshold 
the interactions will be govered by strong annihilation processes. 
In the present study we need also interactions in the 
final $\lcbarsc$, $\scbarsc$, and $\xcbarxc$ systems. Those interactions have been fixed 
by adopting the same philosophy as for $\lcbarlc$ and the parameters are likewise taken 
over from corresponding studies in the strangeness sector \cite{Haidenbauer:1993,Haidenbauer:XX}.
 
%%%%%%%%%%%%%%%%%%%%%%%%%%%%%%%%%%%%%%%%%%%%%%%%%%%%%%%%%%%%%%%
\section{Transition potential from the constituent quark model}
\label{sec:qm}

As an alternative to meson exchange we consider a charm-production potential inspired by quark-gluon 
dynamics. The strange-hadron production in $p \bar p$ reactions has been studied
extensively within the constituent quark model in the past. The best known works are perhaps
those of Kohno and Weise~\cite{Kohno}, Furui and Faessler~\cite{Furui}, 
Burkardt and Dillig~\cite{Burkardt}, Roberts~\cite{Roberts} and Alberg, 
Henley and Wilets~\cite{Alberg}. 
For an extensive list of references see the review \cite{PS185} and for a 
fairly recent work Ref.~\cite{Ortega:2012}. In the present study we 
adopt the interaction proposed by Kohno and Weise, derived in the so-called $^3S_1$ mechanism.
In this model the $\bar ss$ (or $\bar cc$) pair in the final state is created from an initial $\bar uu $ or 
$\bar dd$ pair via $s$-channel gluon exchange, see Fig.~\ref{fig:diag}. After quark 
degrees-of-freedom are integrated out the potential has the form~\cite{Kohno}:
\beq
{V}^{\pbarp \to \lbarl} (r) = \frac{4}{3}A_1(\alpha,\beta)^{3/2} \frac{4\pi \tilde\alpha}{m_G^2}
\delta_{S1}\left(\frac{3}{4\pi\langle {r}^2\rangle }\right)^{3/2}
\exp \left[-\frac{3}{4}B_1(\alpha,\beta) \frac{r^2}{\langle {r}^2 \rangle}\right] \ ,
\label{QM-pot}
\eeq
\beq
{V}^{\pbarp \to \bar \Lambda\Sigma^0,\bar\Sigma^0\Lambda} (r) 
= -\frac{4}{3\sqrt{3}} A_1(\alpha,\beta)^{3/2} \frac{4\pi \tilde\alpha}{m_G^2}
(\delta_{S0}+\frac{2}{3}\delta_{S1})\, 
\left(\frac{3}{4\pi\langle {r}^2\rangle }\right)^{3/2}
\exp \left[-\frac{3}{4}B_1(\alpha,\beta) \frac{r^2}{\langle {r}^2 \rangle}\right] \ ,
\label{QM-potS}
\eeq
\beq
{V}^{\pbarp \to \sobarso} (r) =  \frac{8}{27} A_1(\alpha,\beta)^{3/2} \frac{4\pi \tilde\alpha}{m_G^2}
(\delta_{S0}+\frac{21}{18}\delta_{S1})\,
\left(\frac{3}{4\pi\langle {r}^2\rangle }\right)^{3/2}
\exp \left[-\frac{3}{4}B_1(\alpha,\beta) \frac{r^2}{\langle {r}^2 \rangle}\right] \ ,
\label{QM-potSS}
\eeq
\beq
{V}^{\pbarp \to \spbarsp} (r) =  2\, {V}^{\pbarp \to \sobarso} (r), \quad \quad
{V}^{\pbarp \to \smbarsm} (r) =  0 \ .
\label{QM-potSp}
\eeq

The corresponding expressions for the transitions to charmed baryons (${\pbarp \to \lcbarlc}$, etc.) are formally identical. 
The quantity ${\tilde\alpha}/{m_G^2}$ in Eqs.~(\ref{QM-pot})-(\ref{QM-potSS})
is an effective (quark-gluon) coupling strength, $\langle {r}^2 \rangle$ is the mean square radius 
associated with the quark distribution in the nucleon and $S$ is the total spin in 
the $\pbarp$ system. The effective coupling strength is practically a free parameter 
that was fixed by a fit to the $\pbarp \to \lbarl$ data \cite{Haidenbauer:1992}.
Contrary to Ref.~\cite{Kohno} and to our initial study \cite{Haidenbauer:2010} 
now we take into account the quark-mass dependence of the intrinsic wave functions of 
the baryons. This dependence is encoded in the functions $A_1(\alpha,\beta)$ and 
$B_1(\alpha,\beta)$ for which explicit expressions can be found in Appendix~\ref{app:qm}, 
together with the transition potentials to other channels such as $\xbarx$. For equal quark 
masses $A_1$ and $B_1$ reduce to $1$ so that one recovers the result of Kohno and Weise. 
However, considering the difference in the constituent quark masses of the strange and
the charmed quark one arrives at somewhat different strengths and ranges for the
transition potentials in the strangeness and charm sectors. Choosing 
$\langle {r}^2 \rangle^{1/2}= 0.571$ fm and ${\tilde\alpha}/{m_G^2}=0.252$ fm$^2$ ensures
agreement with the parameters used in our studies of $\pbarp \to \lbarl$ \cite{Haidenbauer:1992}
and $\pbarp \to \lbars$ \cite{Haidenbauer:1993}.

The effective coupling strength depends explicitly on the effective gluon 
propagator $m_G^2$, i.e. on the square of the energy transfer from initial to final quark pair,
cf. Refs.~\cite{Furui,Burkardt,Kohno1}. Heuristically this energy transfer corresponds
roughly to the masses of the produced constituent quarks, i.e. $m_G \approx 2m_q$  
so that we expect the effective coupling strength ${\tilde\alpha}/{m_G^2}$ for charm 
production to be reduced by the ratio of the constituent quark masses of the strange and
the charmed quark squared, $(m_s / m_c)^2 \approx$ (550 MeV / 1600 MeV)$^2$
$\approx$ 1/9 as compared to the one for $\pbarp \to \lbarl$. This reduction factor
is adopted in our calculation for the charm sector. 

In the calculation for the quark-model transition potential the same diagonal 
interactions ($\nbarn\to\nbarn$, $\lbarl\to \lbarl$, ...) as described in the 
last section are employed.
However, the parameters in the optical potentials for $\lbarl$ (cf. Eq.~(\ref{Vaa-ll}))
have been re-adjusted in order to ensure a reproduction of the $\pbarp \to \lbarl$
data \cite{Haidenbauer:1992} and the same has been done in Ref.~\cite{Haidenbauer:1993}
for $\lbars$+c.c. and now for the new data on the $\sbars$ channels. 
For the extension to the charm sector we assume again that 
the $\ycbaryc$ interactions are the same as those for $\ybary$.

\section{Results}
\label{sec:res}

\begin{figure}[t]
\begin{center}
\includegraphics[height=80mm,angle=-90]{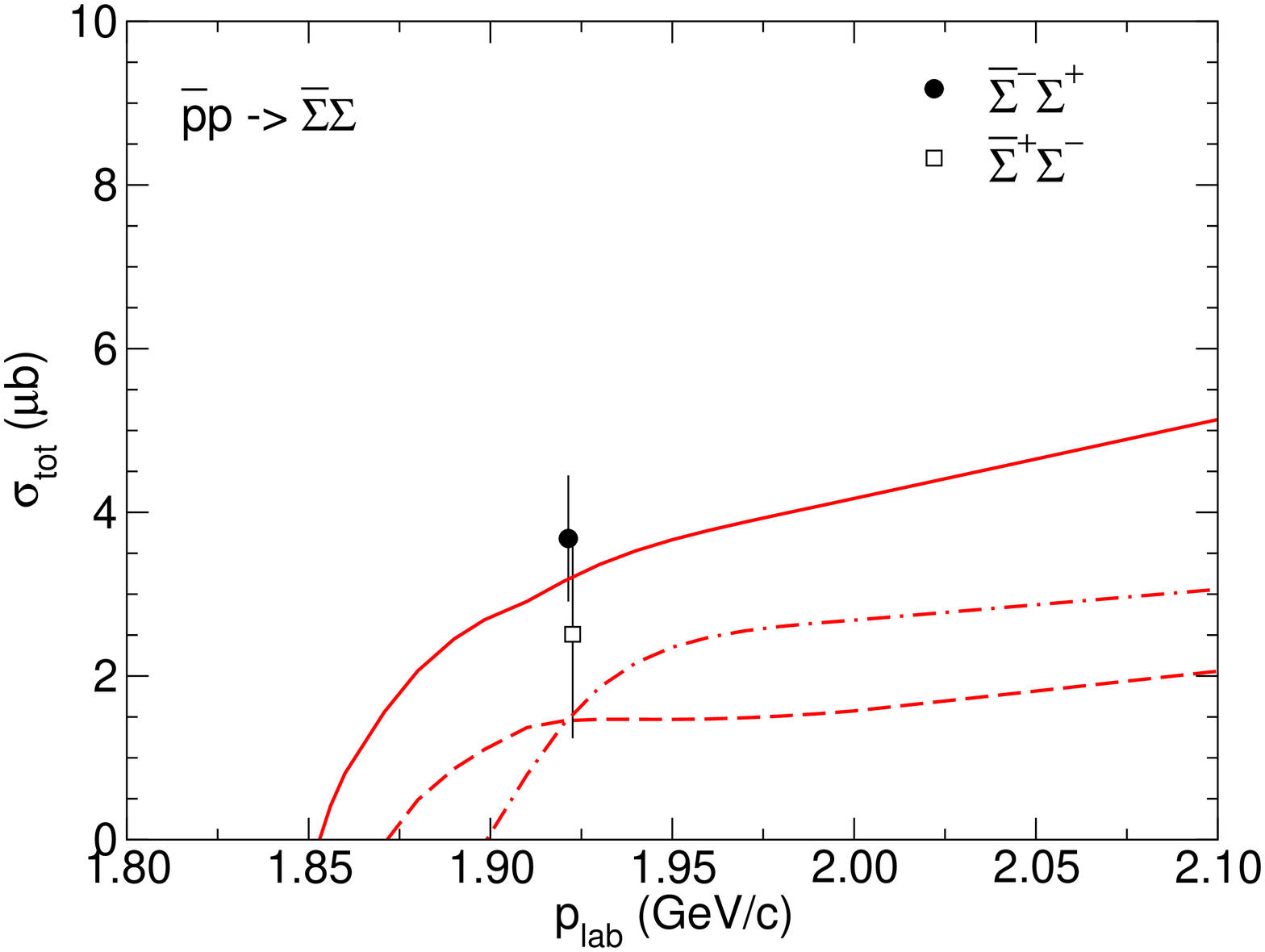}
\includegraphics[height=80mm,angle=-90]{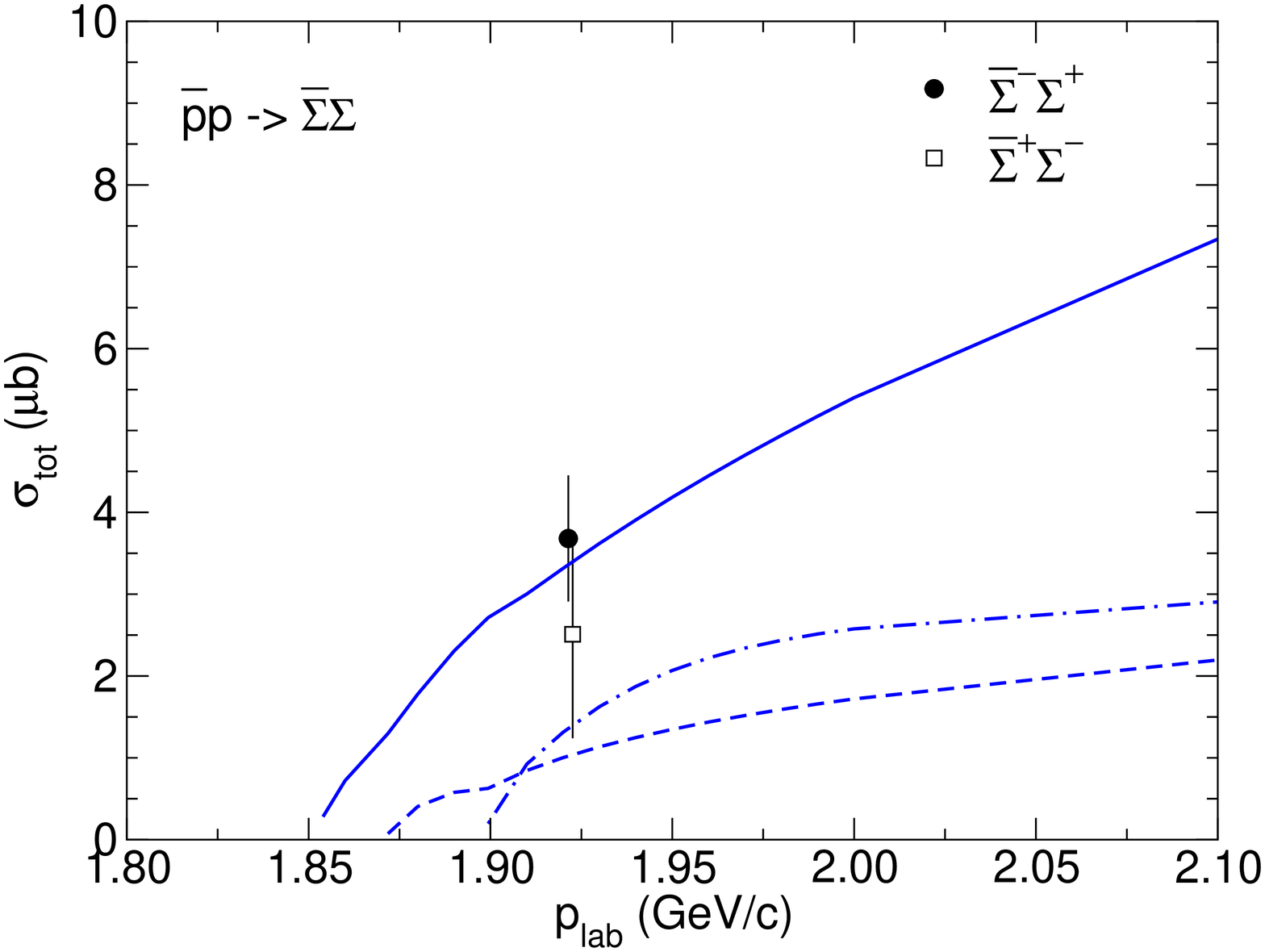}
\caption{Cross sections for $\pbarp \to \Sigma\bar \Sigma$.
On the left results based on the meson-exchange transition potential are
displayed while on the right those for the quark model are shown. 
The solid, dashed, and dash-dotted lines correspond to 
$\spbarsp$, $\sobarso$, and $\smbarsm$, respectively. 
Data taken at $p_{lab}=1.922$~GeV/c are from Refs.~\cite{Barnes:1997,Johansson:1999}. 
The symbols are placed at slightly lower and higher momenta, respectively, so
that the error bars do not overlap. 
}
\label{SSbar}
\end{center}
\end{figure}

Before we present our results for charm production let us discuss
briefly the reaction $\pbarp \to \sbars$. When the J\"ulich group
published their results back in 1993 \cite{Haidenbauer:1993} the
only experimental information on the $\sbars$ channel at low energies 
consisted in a preliminary data point for $\spbarsp$. In the meantime the 
final result for $\spbarsp$ has become available \cite{Barnes:1997} and 
also a measurement for $\smbarsm$ \cite{Johansson:1999}. The latter channel 
is of particular interest because it requires a double charge-exchange
and, therefore, at least a two-step process. In our model calculation
such processes are generated automatically by solving the LS equation (\ref{Tmat}),
and it had been predicted in Ref.~\cite{Haidenbauer:1993} that $\smbarsm$ 
production is by no means suppressed at low energies as one could have expected. 
The actual measurement of the cross section, published several years
after our calculation \cite{Johansson:1999}, nicely confirmed this
prediction, see Fig.~\ref{SSbar} (left).
Results for $\pbarp \to \sbars$ based on the constituent quark-model had 
not been presented in Ref.~\cite{Haidenbauer:1993}. This is done here
for the first time, see Fig.~\ref{SSbar} (right). 
Information on the model results for $\pbarp \to \lbarl$ and $\pbarp \to \lbars$
can be readily found in Refs. \cite{Haidenbauer:1991,Haidenbauer:1992,Haidenbauer:1993}
(for the meson-exchange and the quark-model) and we refrain from reproducing those here. 

\begin{figure}[h]
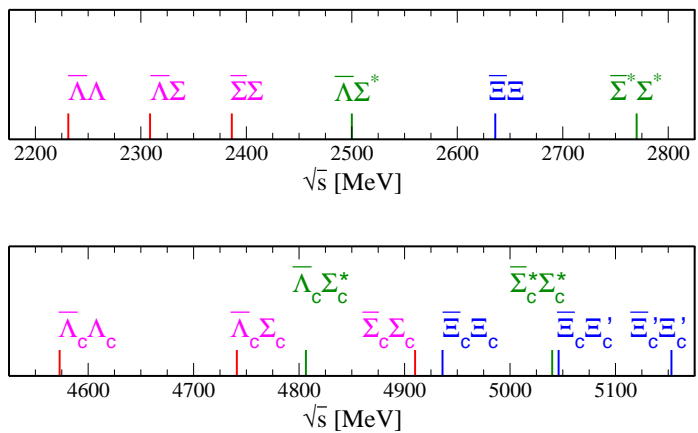

\begin{center}
\vskip 0.3cm
\includegraphics[height=25mm]{kinl.eps}

\vskip 0.6cm
\includegraphics[height=25mm]{kinlc.eps}
\caption{Thresholds of various channels in the strangeness and
charm sectors. $\Sigma^*$ stands for the $\Sigma^*$(1385) and 
$\Sigma_c^*$ for the $\Sigma_c^*$(2520) $3/2^+$ resonances.
Thresholds involving $1/2^-$ baryons such as the $\Lambda$(1405),
for example, are not displayed. 
}
\label{Kinem}
\end{center}
\end{figure}

It is instructive to recall the kinematical situation for the production 
of strange and charmed baryons in $\pbarp$ collisions. This is done in
Fig.~\ref{Kinem} where the thresholds of the various channels are
indicated. One can see that the openings of the 
$\lbarl$, $\lbars$, and $\sbars$ channels are much closer together 
than those of their charmed counterparts. On the other hand, the $\xbarx$
threshold is much farther away than that of $\xcbarxc$. And in the
charmed case there are in addition thresholds involving the $\Xi_c'$.
We indicate also the thresholds of channels that involve the
$3/2^+$ baryons $\Sigma^*$(1385) and $\Sigma_c^*$(2520). 
Those channels are not included in the present study which aims at 
a rough and qualitative estimation of the (strangeness and) charm production 
cross section. It should be said, however, that their presence could have
a sizable quantitative impact on the production cross sections, specifically 
in those reactions whose thresholds lie above the ones for the production
of $3/2^+$ baryons. 

\begin{figure}[t]
\begin{center}
\includegraphics[height=80mm,angle=-90]{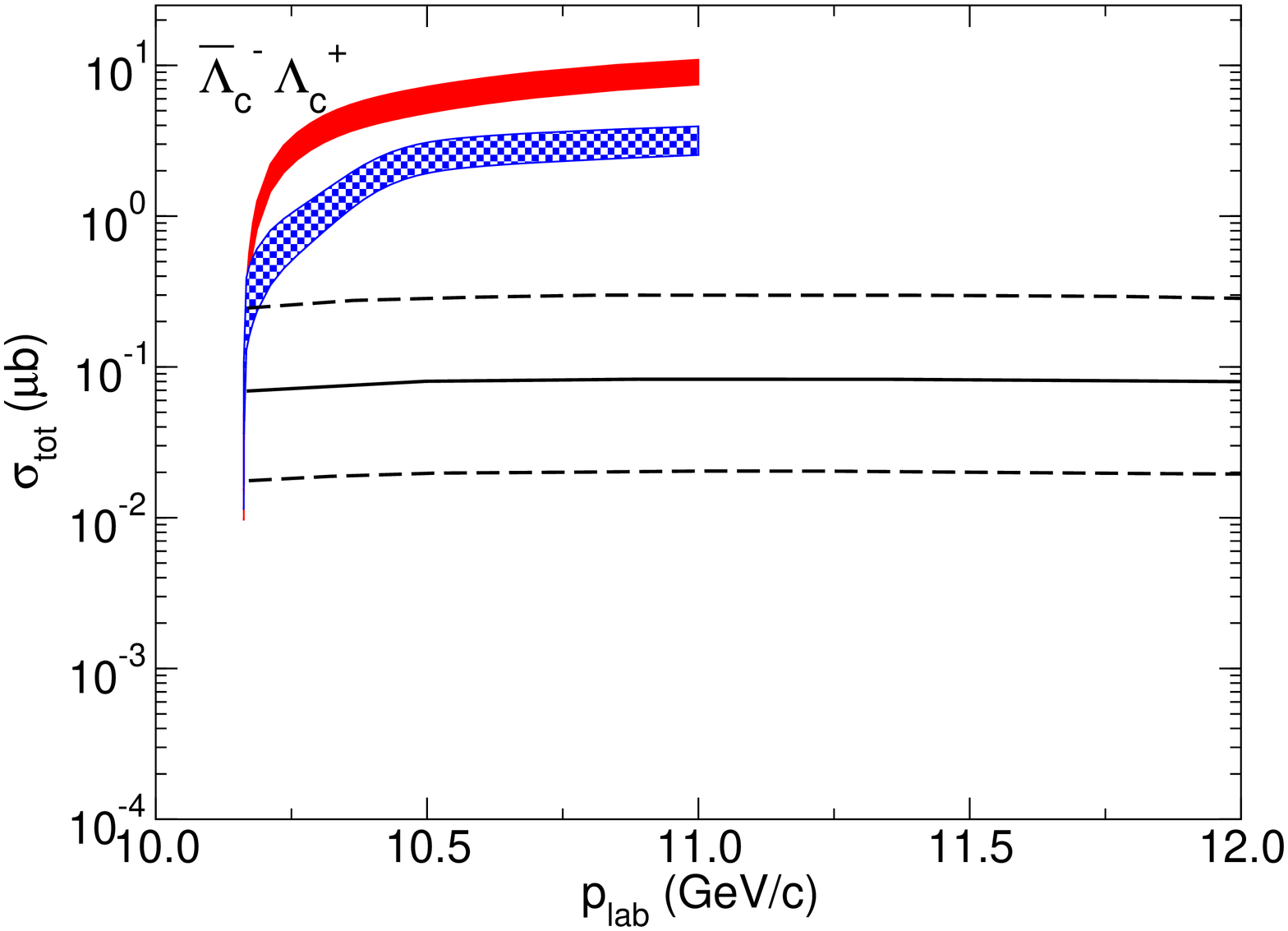}
\includegraphics[height=80mm,angle=-90]{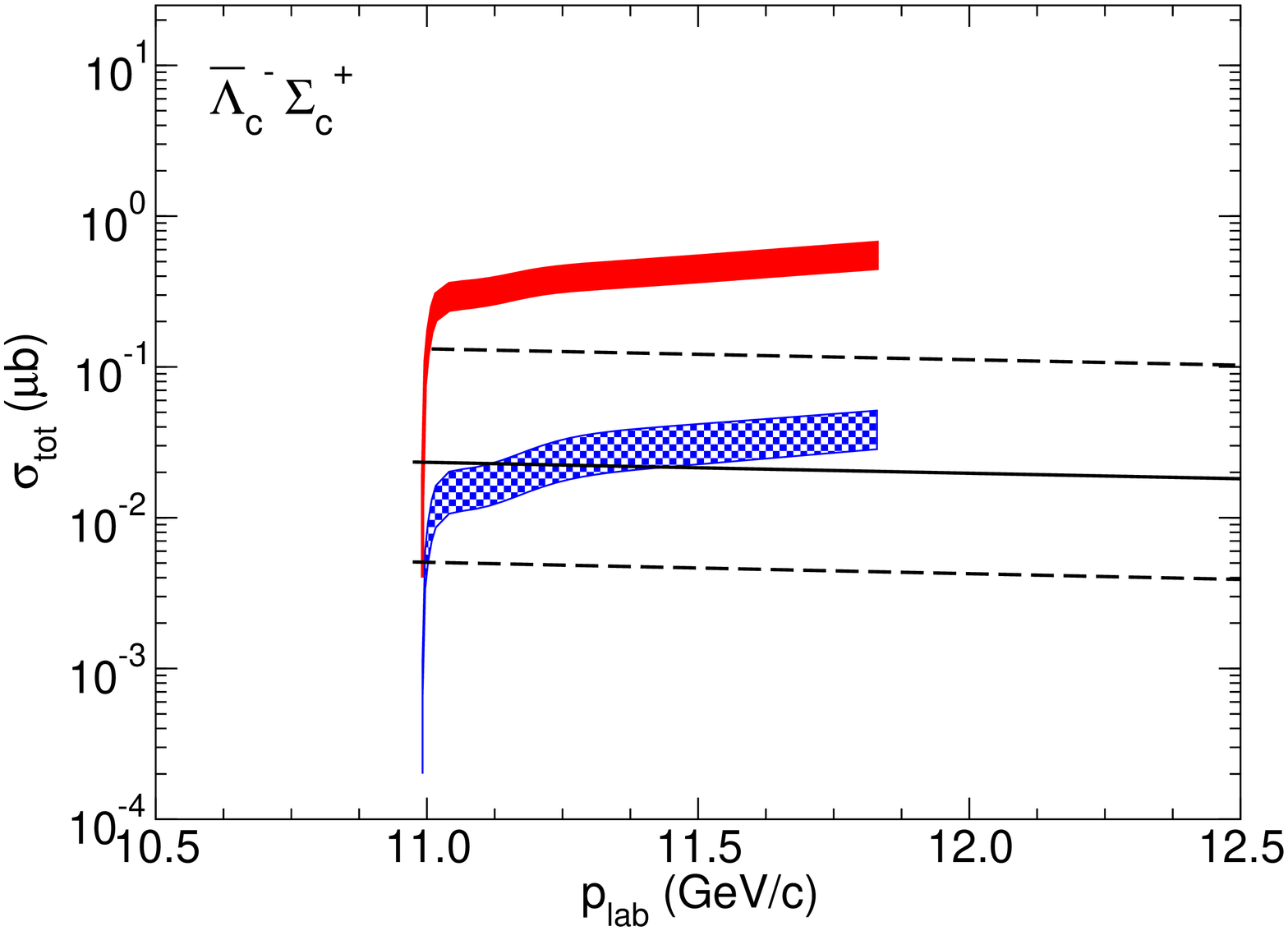}
\caption{
Cross sections for $\pbarp\to \lcbarlc$ (left)
and $\pbarp \to \lcbarsc$ (right) as a function of $p_{lab}$. 
The solid (red) bands are results based on the meson-exchange
transition potential, the hatched (blue) bands are for the
quark model. 
The solid and dashed lines are results taken from Ref.~\cite{Khodjamirian:2012}, 
see text. 
}
\label{LcScbar}
\end{center}
\end{figure}

Predictions for the charm production reactions $\pbarp \to \lcbarlc$ 
and $\pbarp \to \lcbarsc$ are shown in Fig.~\ref{LcScbar}. The meson-exchange 
result for $\pbarp \to \lcbarlc$ is identical to the one presented 
in Ref.~\cite{Haidenbauer:2010}. However, as already said in Sec.~\ref{sec:qm} we no 
longer consider the unrealistic $\nbarn$ model~D because it predicts a too 
large $\pbarp$ cross section and, as a consequence, leads to a much stronger 
reduction of the $\pbarp \to \lcbarlc$ amplitude as compared to the other $\nbarn$ 
potentials (A-C) that reproduce the $\nbarn$ data in the relevant energy 
region very well \cite{Haidenbauer:2010,Haidenbauer:2014}.
Accordingly, the variation of the predicted production cross section due 
to differences in the employed $\nbarn$ ISI, represented by bands 
in Fig.~\ref{LcScbar}, is now much smaller, namely less than a factor 2.
Thus, for $\nbarn$ potentials that not only reproduce the
integrated cross sections but also describe the $\pbarp$ differential 
cross in forward direction satisfactorily the resulting uncertainty in 
the predicted charm production cross sections remains modest. 
Evidently, now the bands from the meson-exchange 
and quark-model transition potentials are clearly separated. 
Note that for the latter in the present work the dependence of the mean 
square radius $\langle r^2 \rangle$ on the quark masses is taken into account,
cf. Sec.~III, and because of that the predictions are slightly
increased as compared to the ones shown in Ref.~\cite{Haidenbauer:2010}.

\begin{table}[h]
\renewcommand{\arraystretch}{1.2}
\centering
\caption{\label{cr25} Production cross sections for strange and charmed
baryons at the excess energy $\varepsilon$ = 25 MeV in $\mu b$. 
The corresponding laboratory momenta are indicated in the table. 
The variations in the charm case are those due to the $\bar NN$ models A-C. 
Note that the results for $\Xi_c'\bar\Xi_c'$ are 
from a truncated coupled-channel calculation, see text.
}
\vskip 0.2cm 
\begin{center}
\begin{ruledtabular}
\begin{tabular}{l||c|cc||c|cc}
%\hline
        & \multicolumn{3}{c||}{Strangeness} & \multicolumn{3}{c}{Charm} \\
\hline
        & $p_{lab}$ & Meson & Quark& $p_{lab}$ & Meson& Quark\\
        & \ (GeV/c) \ & \ exchange \ & model & \ (GeV/c) \ & \ exchange \ & model \\
\hline
$\pbarp \to \lbarl$ ($\lcbarlc$)    & 1.507& 24.7& 27.7& 10.28& 2.65-4.00 & 0.66-1.27\\
$\pbarp \to \lbars$ ($\lcbarsc$)    & 1.724& 5.84& 6.38& 11.12& 0.32-0.49 & 0.01-0.02\\
$\pbarp \to \spbarsp$ ($\scpbarscp$)& 1.942& 3.51& 3.67& 11.98& 0.63-1.09 & 0.001\\
$\pbarp \to \sobarso$ ($\scobarsco$)& 1.942& 1.40& 1.45& 11.98& 0.19-0.29 & 0.001\\
$\pbarp \to \smbarsm$ ($\scmbarscm$)& 1.942& 2.65& 2.86& 11.98& 0.26-0.40 & 0.001\\
$\pbarp \to \xobarxo$ ($\xcpbarxcp$)& 2.677& 0.21& 0.45& 12.15& 0.42-0.60 & 0.003-0.005\\
$\pbarp \to \xmbarxm$ ($\xcobarxco$)& 2.677& 0.17& 0.32& 12.15& 0.17-0.26 & 0.003-0.005\\
$\pbarp \to \xcpbarxcpp$            &      &     &     & 13.32& 0.15-0.22 & (0.5-0.7)$\cdot 10^{-4}$ \ \\
$\pbarp \to \xcobarxcop$            &      &     &     & 13.32& 0.05-0.08 & (0.5-0.7)$\cdot 10^{-4}$ \ \\
%\hline
\end{tabular}
\end{ruledtabular}
\end{center}
\end{table}
\renewcommand{\arraystretch}{1.0}

In order to facilitate a quantitative comparison between the two 
model approaches, but also between the predictions for charm production
with those for the strangeness sector, we provide tables with results 
corresponding to the excess energies of $25$~MeV (Table~\ref{cr25}) and $100$~MeV 
(Table~\ref{cr100}) in the respective channels. One can see from those tables that 
the quark model yields $\pbarp \to \lcbarlc$ cross sections that are roughly a 
factor $2-3$ smaller than the ones based on meson exchange.  

The $\pbarp \to \lcbarsc$ ($\bar \Sigma_c^-\Lambda_c^+$) cross sections 
predicted by the meson-exchange model are more or less an order of magnitude 
smaller than those for $\pbarp \to \lcbarlc$, cf. Fig.~\ref{LcScbar} and 
Tables~\ref{cr25} and \ref{cr100}. This could be somehow expected based 
on the corresponding ratio in the strangeness sector. 
On the other hand, the predictions based on the quark model are much smaller. 
In particular, they are roughly a factor $100$ smaller than the pertinent results
for the $\lcbarlc$ channel, and they are a factor $30$ smaller than the
$\lcbarlc$ results in the meson-exchange picture. 

For the ease of comparison we include in Fig.~\ref{LcScbar} also results 
from Khodjamirian et al.~\cite{Khodjamirian:2012} (solid curve; the dashed 
curves indicate the uncertainty.)
In that study, following Kaidalov and Volkovitsky~\cite{Kaidalov:1994mda},
a non-perturbative quark-gluon string model is used where, however, now
baryon-meson coupling constants from QCD lightcome sum rules are employed.
Interestingly, those results obtained in a rather different framework are more 
or less in line with our quark-model predictions. 

\begin{figure}[t]
\begin{center}
\includegraphics[height=80mm,angle=-90]{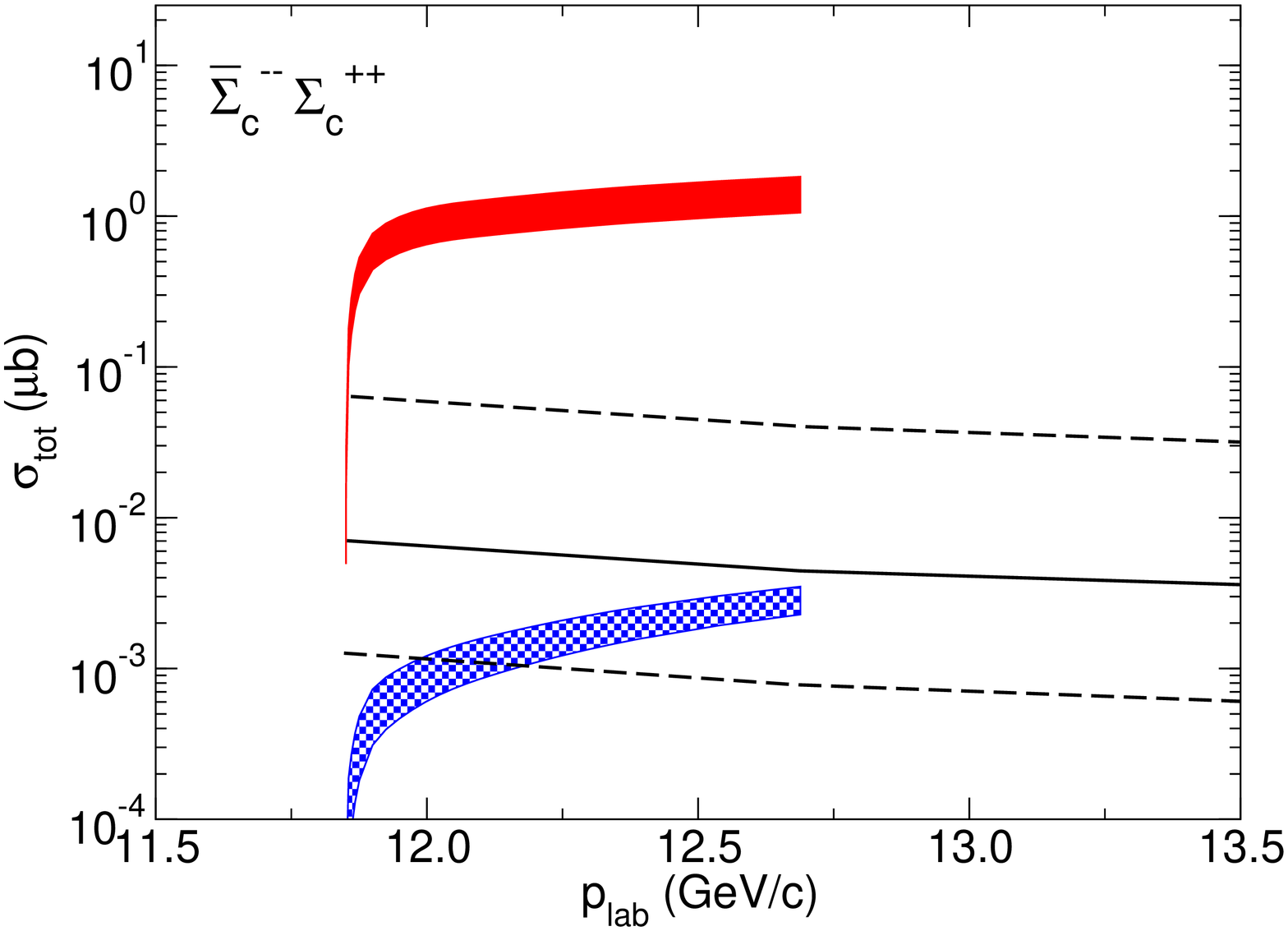}
\includegraphics[height=80mm,angle=-90]{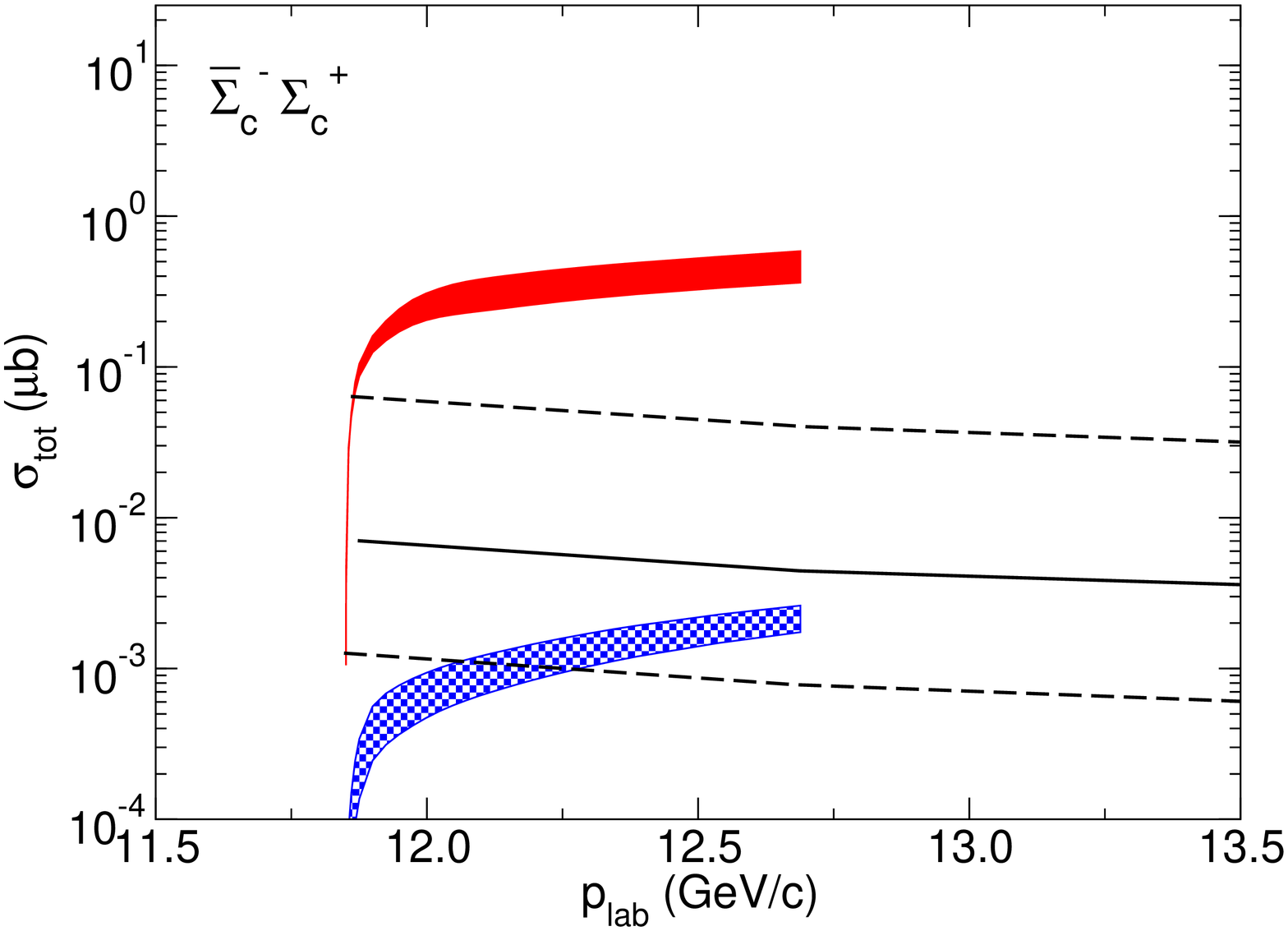}

\includegraphics[height=80mm,angle=-90]{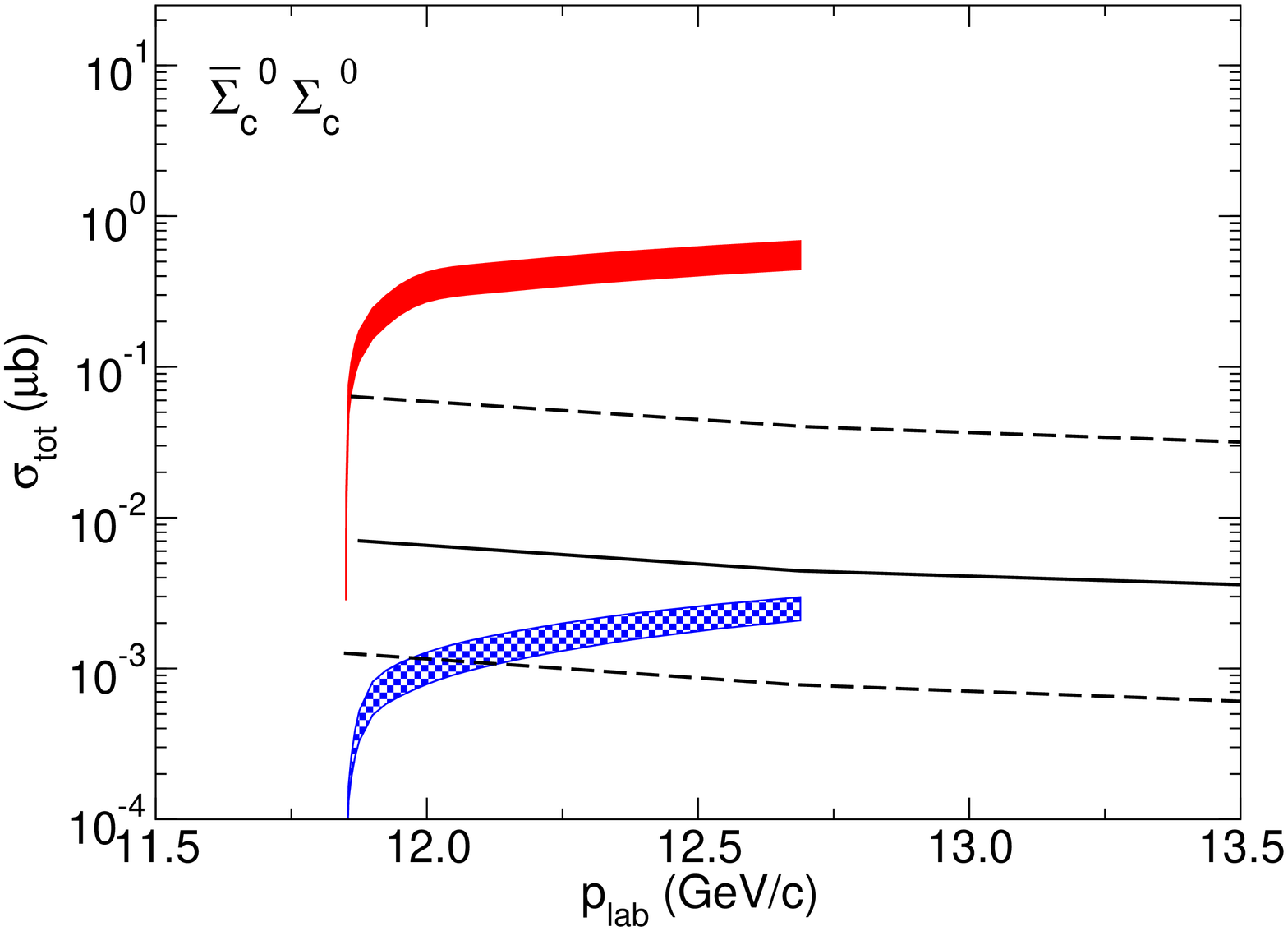}
\caption{Cross sections for $\pbarp \to \scbarsc$ as a function of $p_{lab}$.
Top left: $\scpbarscp$, top right $\scobarsco$, bottom: $\scmbarscm$. \\ 
Same description of curves as in Fig.~\ref{LcScbar}.
}
\label{ScScbar}
\end{center}
\end{figure}

Results for the $\pbarp \to \scbarsc$ channels are presented in 
Fig.~\ref{ScScbar}. The cross sections predicted by the meson-exchange model
are all of similar magnitude, even the one for $ \scmbarscm$
where a two-step process is required. The magnitude is also comparable to the 
cross section for $\pbarp \to \lcbarsc$. Also here the results based on the quark 
model are significantly smaller, i.e. even by roughly three orders of magnitude. 
Again we include here the results from Khodjamirian et al.~\cite{Khodjamirian:2012}.
In this case only isospin averages results are available. 
Let us mention that Kroll et al.~\cite{Kroll:1988cd} have already published 
integrated cross sections for $ \scpbarscp$ more than two decades ago. Their
predictions amount to about $10^{-3}$ $\mu b$ at $p_{lab}=16$ GeV/c and, thus, 
are more or less compatible with those by Khodjamirian et al.

\begin{figure}[t]
\begin{center}
\includegraphics[height=80mm,angle=-90]{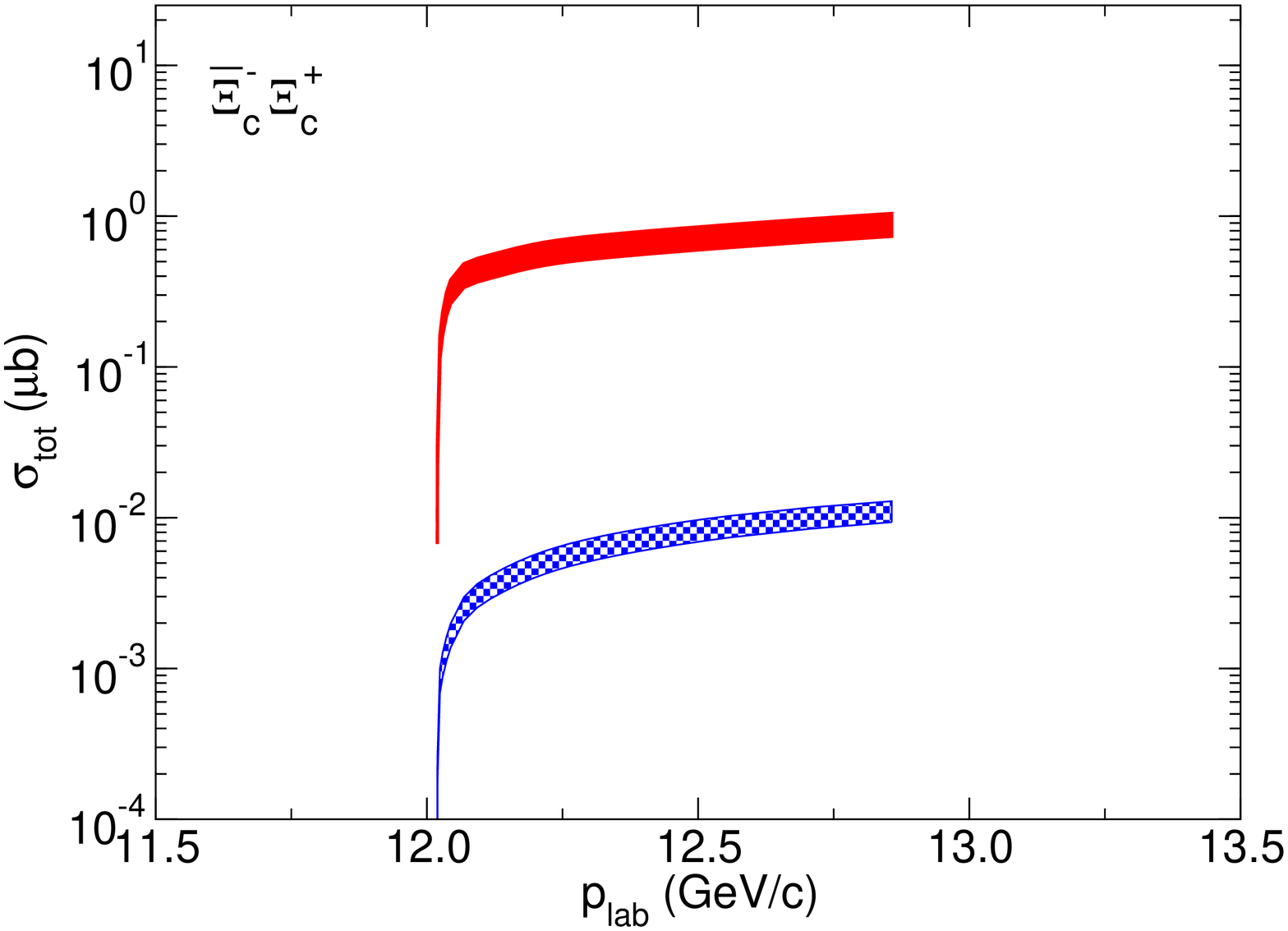}
\includegraphics[height=80mm,angle=-90]{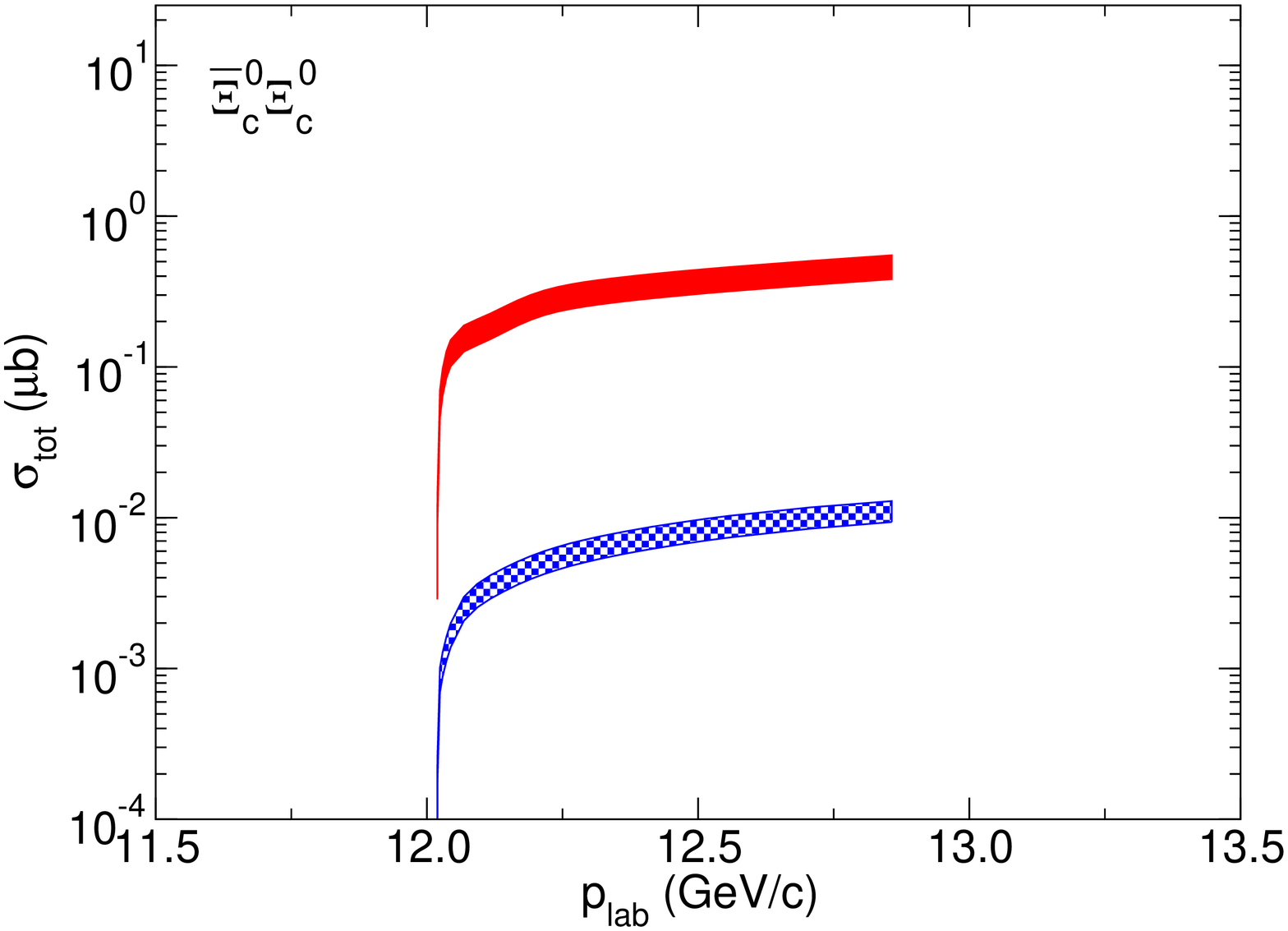}
\caption{Cross sections for $\Xi_c\bar\Xi_c$ as a function of $p_{lab}$.
Left: $\xcpbarxcp$, right $\xcobarxco$.
Same description of curves as in Fig.~\ref{LcScbar}.
}
\label{XXbar}
\end{center}
\end{figure}

Production cross sections for $\pbarp \to \Xi_c\bar\Xi_c$ are displayed in Fig.~\ref{XXbar}. 
The results exhibit a similar pattern to what we already observed for the $\scbarsc$ case. 
Once again the cross sections based on the meson-exchange transition potential
are in the order of $0.1-1$ $\mu b$ while the predictions for the quark model
are orders of magnitude smaller. 
We performed also exploratory calculations for the reaction $\pbarp \to \Xi_c'\bar\Xi_c'$.
Its threshold lies significantly higher than those of the other charmed 
baryons and several more channels are already open, see Fig.~\ref{Kinem}. 
Therefore, in this case only a very rough estimate can be expected from our model
study. Because of that we omitted the $\Xi_c\bar\Xi_c$ and $\Xi_c\bar\Xi_c',\, 
\Xi_c'\bar\Xi_c$ channels in that calculation for simplicity reasons.
The corresponding results are quoted in Tables~\ref{cr25} and \ref{cr100}.

There is a clear trend that the cross sections in the quark model become more
and more suppressed as compared to those from meson-exchange for channels with
higher lying thresholds. 
The main reason for the stronger suppression is presumably related to the exponential $r$ 
dependence of the potential, see the expressions in Sec.~\ref{sec:qm} and Appendix~\ref{app:qm}. 
It amounts to $V^{\mu\nu} (q) \propto {\rm exp}(- q^2\langle r^2 \rangle / 3)$ in 
momentum space with $\bq= \bp_\mu -\bp_\nu$ being the transferred momentum.
With increasing masses of the baryons there is an increasing momentum
mismatch between the on-shell momenta in the initial ($\nbarn$)
and final states and, because of the exponential $q^2$ dependence, the on-shell
matrix elements are strongly reduced for transitions to higher channels.  
In the meson-exchange picture the potential is given by 
$V^{\mu\nu}(q) \propto 1/(-q^2-m^2_M)$, see Eq.~(\ref{Vtrans}). Since $m_M$
is already of the order of $2$~GeV variations in $q^2$ due to the different
charm thresholds have only a moderate effect on the strength of the (on-shell)
potential.

In addition also the interactions in the final $\ycbaryc$ states play
a more important role. For $\lcbarlc$ production we had found that the results 
are rather insensitive to the FSI \cite{Haidenbauer:2010}, leading to a
reduction of the cross section in the order of only 10-15 \% when it is switched 
off altogether. This is no longer the case for channels with higher lying 
thresholds. Indeed, an increased sensitivity is not too surprising in 
view of the fact that some channels like $\scmbarscm$ 
and, of course, $\xcbarxc$ can only be reached by two-step processes,
which means via $\ycbaryc$ FSI effects. We explored the sensitivity by
(arbitrarily) increasing the annihilation in the $\scbarsc$ channel by multiplying 
the strength parameters of the $\scbarsc$ annihilation potential with a factor 2
and found that this reduces the pertinent charm production cross sections by one 
order of magnitude.
Note that specifically for the quark model, where the on-shell transition matrix 
elements are rather small as discussed above, off-shell rescattering in the various 
transitions becomes very important. 

\begin{table}[h]
\renewcommand{\arraystretch}{1.2}
\centering
\caption{\label{cr100} Production cross sections for strange and charmed
baryons at the excess energy $\varepsilon$ = 100 MeV in $\mu b$. 
The corresponding laboratory momenta are indicated in the table. 
The variations in the charm case are those due to the $\bar NN$ models A-C. 
Note that the results for $\Xi_c'\bar\Xi_c'$ are 
from a truncated coupled-channel calculation, see text.
}
\vskip 0.2cm 
\begin{center}
\begin{ruledtabular}
\begin{tabular}{l||c|cc||c|cc}
%\hline
        & \multicolumn{3}{c||}{Strangeness} & \multicolumn{3}{c}{Charm} \\
\hline
        & $p_{lab}$ & Meson & Quark& $p_{lab}$ & Meson& Quark\\
        & \ (GeV/c) \ & \ exchange \ & model & \ (GeV/c) \ & \ exchange \ & model \\
\hline
$\pbarp \to \lbarl$ ($\lcbarlc$)    & 1.719& 72.6& 70.6& 10.66& 5.65-8.37 & 2.22-3.49\\
$\pbarp \to \lbars$ ($\lcbarsc$)    & 1.937& 10.6&  9.5& 11.50& 0.60-0.91 & 0.02-0.04\\
$\pbarp \to \spbarsp$ ($\scpbarscp$)& 2.157& 5.63& 8.48& 12.38& 0.91-1.58 & 0.002\\
$\pbarp \to \sobarso$ ($\scobarsco$)& 2.157& 2.35& 2.77& 12.38& 0.30-0.46 & 0.002\\
$\pbarp \to \smbarsm$ ($\scmbarscm$)& 2.157& 3.27& 3.66& 12.38& 0.38-0.58 & 0.002\\
$\pbarp \to \xobarxo$ ($\xcpbarxcp$)& 2.904& 0.40& 0.94& 12.55& 0.62-0.87 & 0.007-0.010\\
$\pbarp \to \xmbarxm$ ($\xcobarxco$)& 2.904& 0.29& 0.76& 12.55& 0.31-0.45 & 0.007-0.010\\
$\pbarp \to \xcpbarxcpp$            &      &     &     & 13.74& 0.27-0.39 & (0.1-0.2)$\cdot 10^{-3}$ \ \\
$\pbarp \to \xcobarxcop$            &      &     &     & 13.74& 0.08-0.13 & (0.1-0.2)$\cdot 10^{-3}$ \ \\
%\hline
\end{tabular}
\end{ruledtabular}
\end{center}
\end{table}
\renewcommand{\arraystretch}{1.0}

The charm production cross sections based on the meson-exchange picture depend
also sensitively on the form-factor parameters at the $NY_c D$ and $NY_c D^*$ vertices. 
As said in Sec. II, for the results discussed above a cutoff mass of $\Lambda =$ 3 GeV 
has been used. When reducing this value to 2.5 GeV the cross sections for
$\pbarp \to \lcbarlc$ drop by roughly a factor 3 \cite{Haidenbauer:2010}.  
For the other charm production channels considered in the present paper 
such a decrease of the cutoff mass in the transition potential yields 
a reduction of a factor $5$ in the cross sections.
One can view that variation as a further uncertainty of the predictions based 
on the meson-exchange model. If so one can conclude that the results of the
meson-exchange and quark transition potentials for $\pbarp \to \lcbarlc$  
are indeed compatible with each other. However, this is definitely not the
case for the other charm production channels considered. 
In principle, employing even smaller cutoff masses would further decrease 
the cross sections of the meson-exchange charm-production mechanism.  
However, as argued in Ref.~\cite{Haidenbauer:2010}, in view of the 
fact that the exchanged mesons have a mass of around 1.9 to 2 GeV
we consider values below 2.5 GeV as being not really realistic. 

Finally, a comment on the SU(4) flavor symmetry which is used here as guideline for 
providing constraints on the pertinent baryon-meson coupling constants.
As already said in the Introduction, recent calculations of the relevant ($D$ and $D^*$) 
coupling constants within QCD light-cone sum rules suggest that 
the actual deviation from the SU(4) predictions could be in the order of a factor $2$ 
or smaller, see Table~1 in Ref.~\cite{Khodjamirian:2012}. Indeed, in several cases the ratio
of the $NY_cM$ to $NYM$ coupling constants turned out to be practically consistent with 
SU(4) symmetry (where that ratio is $1$) within the quoted uncertainty. In any case, 
since the coupling constants enter quadratically into the potential, see Eq.~(\ref{Vtrans}), 
and with the 4th power into the cross sections it follows that a factor $2$ ($1.5$)
in the coupling constant implies roughly a variation in the order of a factor $16$ ($5$)
in the cross section. Such variations are larger than those from the $\pbarp$ ISI represented
by the bands. However, they are well within the difference we observe between
the predictions based on the meson-exchange transition potential and those of the
quark model. 
 
\section{Summary}
\label{sec:sum}

In this paper we presented predictions for the charm-production
reaction $\pbarp \to \lcbarlc$, $\lcbarsc$, $\scbarsc$, and $\xcbarxc$.
The production process is described within the meson-exchange picture in 
close analogy to our earlier studies on $\pbarp \to \lbarl$ \cite{Haidenbauer:1991},
$\lbars$, $\sbars$ \cite{Haidenbauer:1993}, and $\xbarx$ \cite{Haidenbauer:XX} 
by connecting the dynamics via SU(4) symmetry. 
The calculations were performed within a coupled-channels framework so
that the interaction in the initial $\pbarp$ interaction, which plays a 
crucial role for reliable predictions, can be taken into account
rigorously. The interactions in the various $\ycbaryc$ channels and
the transitions between those channels are also included. 

The obtained $\lcbarsc$ ($\scbarlc$) production cross sections are in the order 
of $0.5-1$ $\mu b$ for energies not too far from the threshold. 
Thus, they are about a factor $10$ smaller than the corresponding cross 
sections for $\lcbarlc$.
The $\scbarsc$ cross sections are likewise in the order of 
$0.5-1$ $\mu b$ where those for $\scpbarscp$ are predicted to be somewhat 
larger than those for $\scobarsco$ and $\scmbarscm$.
The cross sections for $\xcbarxc$ production, for which the threshold is only 
slightly higher than the one for $\scbarsc$, are found to be 
around $0.5$ $\mu b$.

In order to shed light on the model dependence of our results 
we investigated the effect of replacing the meson-exchange transition potential
by a charm-production mechanism derived in a quark model. 
In our earlier work on the reaction $\pbarp \to \lcbarlc$ we had found that 
both pictures lead to predictions of essentially the same order of 
magnitude \cite{Haidenbauer:2010}. Thus, it seemed that the details of the 
production mechanism do not matter, only the involved scales and these are 
fixed by the masses of the exchanged mesons or, correspondingly, the 
constituent masses of the produced charmed quarks.
Now, it turned out that our conclusion drawn from that work was perhaps 
too optimistic. The extension of the study to other charmed baryons in the present 
work revealed drastic differences between the predictions of the
two production mechanisms for channels with higher thresholds. 
Specifically, for $\pbarp \to \lcbarsc$ $(\scbarlc)$ the quark model
yields results that are more than one order of magnitude smaller than
those obtained for the meson-exchange model and in case of 
$\pbarp \to \scbarsc$ or $\pbarp \to \xcbarxc$ the differences even amount 
to three orders of magnitude.

Clearly, this large difference or uncertainty in our predictions is a bit disillusioning. 
But to some extent it does not really come unexpected. While for the lowest
channel, $\lcbarlc$, the magnitude of the cross section is mostly influenced
by the initial $\pbarp$ interaction (which is known and fixed from
experimental data) this is no longer the case for the other reactions.
Here two-step processes of the form $\pbarp \to \lcbarlc \to \scbarsc$, 
$\pbarp \to \lcbarsc\, (\scbarlc) \to\scbarsc$, etc. become increasingly
important. Accordingly, the influence of the interactions in the $\lcbarlc$, 
$\lcbarsc$, ...,  channels become more significant and those are not constrained 
by any empirical information. 
Specifically, for the quark-model results these interactions play a
decisive role because direct transitions are suppressed due to the large
momentum mismatch. 
Accepting the difference between the predictions based on meson 
exchange and those for the quark model as the basic uncertainty of our 
model calculation leaves ample room and, thus, is not so uncouraging 
for pertinent measurements. The results for meson exchange taken alone
convey a much more optimistic perspective for experimental efforts. 
In any case, which of those scenarios is closer to reality can be only 
decided by performing concrete experiments that will hopefully be 
pursued at FAIR in the future. 

\vskip 0.2cm 
\noindent
{\bf Acknowledgments}
We thank Ulf-G. Mei{\ss}ner for his careful reading of our manuscript and 
Albrecht Gillitzer for stimulating discussions. 
Work partially supported by Conselho Nacional de Desenvolvimento Cient\'{\i}fico e Tecnol\'ogico 
- CNPq, Grant No. 305894/2009-9 (GK), and Fun\-da\-\c{c}\~ao de Amparo \`a Pesquisa do Estado de 
S\~ao Paulo - FAPESP, Grant No. 2013/01907-0 (GK).  

\appendix

\section{Quark model}
\label{app:qm}

The microscopic quark-model interaction of the strange- and charm-baryon production 
potentials is inspired by an $s-$channel one-gluon exchange amplitude for light 
quark-antiquark $\bar l l$ pair annihilation and heavy quark-antiquark $\bar h h$ pair 
creation that can be parametrized in terms of an effective quark-gluon coupling strength 
$\tilde \alpha/m^2_G$ as:  
\bea
V(\bar{l}l \rightarrow \bar{h} h) = - \frac{4\pi \tilde \alpha_s}{m^ 2_G} \, C \, 
\bS^2 \, 
\delta^3(\br_l - \br_{\bar l}) \, \delta^3(\br_h - \br_{\bar h}) \, \delta^3(\br_h - \br_l),
\label{Vann}
\eea
where $C$ is the color matrix 
\bea
C(\bar{l}l  \rightarrow \bar{h}h) &=& \frac{1}{6} \sum^8_{a=1} \left[ 
\left(\frac{\lambda^a}{2}\right)_{\bar{l}l} - \left(\frac{\lambda^a}{2}\right)^*_{\bar{h}h}
\right]^2, 
\eea
and $\bS$ is the total spin of the light quark-antiquark pair (or of the heavy antiquark-quark
pair, any quark mass factors involved in either case are absorbed in the effective coupling).
Also, $l$ stands for $(u,d)$ and, depending on the case, $h$ for $s$ or $c$. For example,
while in the process $\bar pp \rightarrow \bar \Lambda \Lambda \, (\bar\Lambda_c\Lambda_c)$, 
$h = s\,(c)$, in $\bar\Lambda\Lambda \rightarrow \bar\Xi_c\Xi_c$, $l=(u,d)$ 
and $h=c$, and so on.

To evaluate the transition potential, we need quark-model wave
functions for the baryons and antibaryons. For simplicity, we use harmonic oscillator 
wave functions, that for the ground states are given by
\beq
\phi_B(\br_1,\br_2,\br_2) = \phi_B(\brho,\blam) = \left(\frac{1}{\pi b^2_\rho}\right)^{3/4} 
\left(\frac{1}{\pi b^2_\lambda}\right)^{3/4}\,
\exp \left(- \frac{1}{2b^2_\rho}\brho^2 - \frac{1}{2b^2_\lambda}\blam^2 \right), 
\label{intrinsic}
\eeq
where $\brho$ and $\blam$ are the Jacobi coordinates $\brho = \br_1 - \br_2$ and 
$\blam = \br_3 - (m_1 \br_1 + m_2 \br_2)/(m_1 + m_2)$, with $m_1$, 
$m_2$ and $m_3$ being the quarks masses. For example, for the proton we have $m_1 = m_2 = m_u$ 
and $m_3 = m_d$ (in the present paper, we take $m_u = m_d \equiv m$),  
$b^p_\rho = \sqrt{ 2 \, \langle r^2 \rangle}$, and $b^p_\lambda = \sqrt{3/2\,\langle r^2 \rangle}$,
where $\langle r^2\rangle$ is the proton mean square radius. For~$\Lambda$~($\Sigma$), $m_1 = m_2 = m$, 
$m_3 = m_s$, and the size parameters $b^\Lambda_\rho$ and $b^\Lambda_\lambda$ are related to 
those of the proton as $b^\Lambda_\rho = b^p_\rho$ and $b^\Lambda_\lambda = \sqrt{\alpha} \,
b^p_\lambda$ (and analogous for $\Sigma$), where $\alpha$ depends on the quark masses as
\beq
\alpha^2 = \frac{m_s + 2m}{3m_s}.
\label{alpha}
\eeq
For $\Lambda_c$~($\Sigma_c$), $m_s$ in Eq.~(\ref{alpha}) is to be replaced by the charmed quark 
mass $m_c$. 
For $\Xi$ one has $m_1 = m_2 = m_s$ and $m_3 = m$ and, analogous to $\Lambda$, one can relate
the respective size parameters to those of the proton as $b^\Xi_\rho = b^p_\rho$ and 
$b^\Xi_\lambda = \sqrt{\beta} \, b^p_\lambda$, with 
\beq
\beta^2 = \frac{m + 2m_s}{3m}.
\label{beta}
\eeq 
Finally, for the $\Xi_c$ and $\Xi^\prime_c$ states, which involve $u(d)$, $s$ and $c$ quarks,
in order to keep the calculation simple we define an average mass $\bar m = (m + m_s)/2$ so
that in the wave function (\ref{intrinsic}) the size parameters are related to those of
the proton as $b^{\Xi_c}_\rho = b^p_\rho$ and $b^{\Xi_c}_\lambda = \sqrt{\bar\beta} \, b^p_\lambda$,
with
\beq
{\bar\beta}^2 = \frac{m_c + 2\bar m}{3m_c}.
\label{beta-bar}
\eeq   

Given the microscopic interaction and the baryon and antibaryon wave functions, one can evaluate
rather easily the transition potentials. For the transitions $\bar pp \rightarrow \bar\Lambda\Lambda,
\bar\Sigma^0\Sigma^0, \bar\Lambda\Sigma^0, \bar\Sigma^-\Sigma^+$ the potentials have been given explicitly 
in Eqs.~(\ref{QM-pot})-(\ref{QM-potSp}), where the functions $A_1(\alpha,\beta)$ and $B_1(\alpha,\beta)$  
take into account the different sizes of the baryons due to quark-mass differences encoded in
the parameters $\alpha$ and $\beta$ defined above:
\beq
A_1(\alpha,\beta) = \frac{2}{1 + \alpha}, \hspace{1.0cm}
B_1(\alpha,\beta) = \frac{1 + \alpha}{2 \alpha}.
\hspace{1.0cm}\label{A1-B1}
\eeq
The expressions for the corresponding charm-production potentials 
$\bar pp \rightarrow \bar\Lambda^-_c\Lambda^+_c$, ..., are identical. But $\alpha$ and $\beta$ differ and,
accordingly, the values of the factors $A_1$ and $B_1$. And, of course, also the effective coupling constant
is different. 
The transition potentials for double-strange baryon production, $\bar\Lambda\Lambda, \bar\Sigma^0\Sigma^0, 
\bar\Lambda\Sigma^0, \bar\Sigma^-\Sigma^0 \rightarrow \bar\Xi^0\Xi^0, \bar\Xi^+\Xi^-$, can be written
generically as
\beq
V^{\text{2-s\,prod}} (r) =  A_2(\alpha,\beta)^{3/2} \left(\frac{4\pi \tilde\alpha}{m^2_G}\right) 
\left(\chi_0 \, \delta_{S0} + \chi_1 \,\delta_{S1}\right)  
\left(\frac{3}{4\pi\langle r^2 \rangle}\right)^{3/2} 
\exp \left[-\frac{3}{4} B_2(\alpha,\beta) \frac{r^2}{\langle r^2 \rangle}\right],
\label{V2s}
\eeq
where
\bea
A_2(\alpha,\beta) &=& \frac{2^9 \, \alpha^4 \, \beta}{\left[3 + 5 \beta + \alpha\left(5 + 3\beta\right) \right]\,
\left[3 + 5\beta + 6\alpha^2\left(-1+\beta\right) + 12 \alpha^3 \beta + \alpha^4 \left(3+9\beta\right) \right] },
\label{Aab}
\\[0.3true cm]
B_2(\alpha,\beta) &=& \frac{2\,\alpha^3 \, \left[3 +  5\beta + \alpha\left(5 + 3\beta\right) \right]}
{3 + 5\beta + 6\alpha^2\left(-1+\beta\right) + 12 \alpha^3 \beta + \alpha^4 \left(3+9\beta\right) }.
\label{Bab}
\eea
and $\chi_0$ and $\chi_1$ are color-spin-flavor coefficients whose values are given in Table~\ref{tab:2s}
(we use the phase conventions of Ref.~\cite{Close} for the spin-flavor wave functions).

\begin{table}[ht]
\caption{Color-spin-flavor factors $\chi_0$ and $\chi_1$ for the transitions to double-strange baryons 
$\bar\Xi^{(0,+)}\Xi^{(0,-)}$.}
\begin{center}
\begin{ruledtabular}
\begin{tabular}{cccccc}
    \multirow{1}{*}{\text{Initial state}$\;\;\;\rightarrow$}
                     & $\bar\Lambda \Lambda$        & $\bar\Sigma^0\Sigma^0$     &  $ \bar\Lambda\Sigma^0$      
                     & $\bar\Sigma^-\Sigma^+$ \;\;\;
                     & {\text{Final State}\;\;$\downarrow$\;\;\;\;}\\
    \hline \\[-0.25cm]
    \multirow{4}{*}{{\vspace{0.75cm}S=0}}
    & $\frac{8}{9}$   & -$\frac{16}{27}$ & $\frac{4}{9\sqrt{3}}$  & -$\frac{32}{27}$ 
    &{$\bar\Xi^0\;\Xi^0$}\\[0.1cm]
    & $\frac{8}{9}$   & -$\frac{16}{27}$ & -$\frac{4}{9\sqrt{3}}$ & $0$ 
    &{$\bar\Xi^+\Xi^-$}  \\[0.2cm]
    \hline \\[-0.25cm] 
    \multirow{4}{*}{{\vspace*{0.75cm}S=1}}
    & $\frac{28}{27}$ & $\frac{52}{81}$  & $\frac{16}{9\sqrt{3}}$ & $\frac{104}{27}$ 
    &{$\bar\Xi^0\;\Xi^0$}\\[0.1cm]
    & $\frac{28}{27}$ & $\frac{52}{81}$  & -$\frac{16}{9\sqrt{3}}$ & $0$ 
    &{$\bar\Xi^+\Xi^-$}\\[0.1cm]
\end{tabular}
\end{ruledtabular}
\end{center}
\label{tab:2s}
\end{table}

In the production of the charmed antibaryon-baryon states $\bar\Xi^{(0,-)}_c\Xi^{(0,+)}_c$ and 
$\bar\Xi^{\prime (0,-)}_c\Xi^{\prime (0,+)}_c$, there are two situations to distinguish, those 
with strange antibaryon-baryon ($\bar\Lambda\Lambda, \bar\Sigma^0\Sigma^0, \bar\Lambda\Sigma^0, 
\bar\Sigma^0\Lambda$) in the initial states and those with charmed antibaryon-baryon 
($\bar\Lambda^-_c\Lambda^+_c, \bar\Sigma^-_c\Sigma^+_c, 
\bar\Lambda^-_c\Sigma^+_c, \bar\Sigma^-_c\Lambda^+_c, \bar\Sigma^{--}_c\Sigma^{++}_c$).
While in the first case a anticharm-charm quark pair is created, in the second a antistrange-strange
quark pair is created and the symmetry of the wave functions leads to different transition 
potentials in the two cases. The corresponding transition potentials are of the generic form given in 
Eq.~(\ref{V2s}), with the coefficients $\chi_0$ and $\chi_1$ given in Table~\ref{tab:charm}, and the 
functions $A_2$ and $B_2$ replaced by 
\beq
A_3(\alpha,\bar\beta) = A_2(\alpha,\bar\beta), \hspace{1.0cm} B_3(\alpha,\bar\beta) = B_2(\alpha,\bar\beta)
\eeq
for the strange antibaryon-baryon initial states and
\bea
A_4(\alpha_c,\bar\beta) &=& \frac{2^4 \, \alpha^4_c \, \beta}{\left(\alpha_c + \bar\beta\right)
\left(3\alpha^4_c + 2\bar\beta + 3 \alpha^3_c \bar\beta\right)},
\label{Caa}
\\[0.5true cm]
B_4(\alpha_c,\bar\beta) &=& \frac{4\,\alpha^3_c \, \left(\alpha_c + \bar\beta\right)}
{3\alpha^4_c + 2 \bar\beta + 3\alpha^3_c\bar\beta}
\label{Daa}
\eea
for the charmed antibaryon-baryon initial states, with $\alpha_c$ being the charmed counterpart 
of $\alpha$:
\beq
\alpha^2_c = \frac{m_c + 2m_u}{3m_c}.
\label{alpha-c}
\eeq
Though we provide here all transition potentials between the strangeness and the charm sectors 
for completeness reasons, it should be said that only transitions of the form
$\pbarp \to \lcbarlc \to \xcbarxc$, etc., are included in the actual coupled-channel calculation. 
Transitions via strange baryons like $\pbarp \to \lbarl \to \xcbarxc$ are ignored. 
We expect such processes to be less significant. At least, in our study of the production of 
the charm-strange meson $D_s$ in $\pbarp \to D^+_sD^-_s$ it had turned out that two-step processes 
involving strange hadrons are practically negligible \cite{Haidenbauer:2014}. 
 
\begin{table}[ht]
\caption{Color-spin-flavor factors $\chi_0$ and $\chi_1$ for the transitions to $\bar\Xi^{(0,-)}_c\Xi^{(0,+)}_c$ and
$\bar\Xi^{\prime (0,-)}_c\Xi^{\prime (0,+)}_c$ final states}
\begin{center}
\begin{ruledtabular}
\begin{tabular}{ccccccccc}
    \multirow{1}{*}{\text{Initial state}$\;\;\;\rightarrow$}
                     & $\bar\Lambda \Lambda$        & $\bar\Sigma^0\Sigma^0$     &  $ \bar\Lambda\Sigma^0$      
                     & $\bar\Lambda^-_c\Lambda^+_c$ & $\bar\Sigma^-_c\Sigma^-_c$ &  $ \bar\Lambda^-_c\Sigma^+_c$ 
                     & $\bar\Sigma^{--}_c\Sigma^{++}_c$ \;\;\;
                     & {\text{Final State}\;\;$\downarrow$\;\;\;\;}\\
    \hline \\[-0.25cm]
    \multirow{4}{*}{{\vspace{-0.5cm}S=0}}
    & -$\frac{2}{9}$    & $0$                     & $0$                      & $\frac{2}{3}$  
                        & -$\frac{2}{9}$     & $0$                     & $0$
    &{$\bar\Xi^0_c\;\Xi^0_c$}\\[0.1cm]
                        &   $0$              & $0$                     & $0$                      & $\frac{2}{3}$  
                        & -$\frac{2}{9}$     & $0$                     & -$\frac{4}{9}$                
    &{$\bar\Xi^-_c\Xi^+_c$} \\[0.1cm]
                        & $\frac{4}{9}$      & $\frac{4}{9}$           & $\frac{11}{27}$          & -$\frac{2}{9}$ 
                        & $\frac{10}{27}$    & $\frac{4}{9\sqrt{3}}$   & $0$
    &{$\bar\Xi^{\prime 0}_c\Xi^{\prime 0}_c$} \\[0.1cm]
                        & $\frac{4}{9}$      & $\frac{4}{27}$          & -$\frac{4}{9\sqrt{3}}$  & -$\frac{2}{9}$ 
                        & $\frac{10}{27}$    & -$\frac{4}{9\sqrt{3}} $ & $\frac{20}{27}$
    &{$\bar\Xi^{\prime -}_c\Xi^{\prime +}_c$} \\[0.2cm]
    \hline \\[-0.25cm] 
    \multirow{4}{*}{{\vspace{-0.5cm}S=1}}
      & $\frac{2}{27}$     & $\frac{2}{3}$           & -$\frac{2}{3\sqrt{3}}$   & $\frac{2}{3}$  
                        & $\frac{2}{27}$     & $0$                     & $0$
      &{$\bar\Xi^0_c\;\Xi^0_c$}\\[0.1cm]
                        & $\frac{2}{9}$      & $\frac{2}{3}$           & -$\frac{2}{3\sqrt{3}}$   & $\frac{2}{3}$  
                        & $\frac{2}{27}$     & $0$                     & $\frac{4}{27}$
      &{$\bar\Xi^-_c\Xi^+_c$}\\[0.1cm]
                        & $\frac{14}{27}$    & $\frac{14}{27}$         & $\frac{14}{27}$          & $\frac{2}{27}$ 
                        & $\frac{62}{81}$    & -$\frac{4}{27\sqrt{3}}$ & $0$
      &{$\bar\Xi^{\prime 0}_c\Xi^{\prime 0}_c$}\\[0.1cm]
                        & $\frac{14}{27}$    & $\frac{14}{81}$         & -$\frac{14}{27\sqrt{3}}$ & $\frac{2}{27}$ 
                        & $\frac{62}{81}$    & $\frac{4}{27\sqrt{3}}$  & $\frac{124}{81}$
      &{$\bar\Xi^{\prime -}_c\Xi^{\prime +}_c$}\\[0.1cm]
\end{tabular}
\end{ruledtabular}
\end{center}
\label{tab:charm}
\end{table}

\section{SU(4) considerations}
\label{app:su4}

For calculating the baryon-baryon-meson coupling constants within
the assumed SU(4) symmetry we utilize here the tensors
$\psi_{\mu\nu\lambda}$ ($\mu,\nu,\lambda = 1, ..., 4$) introduced
by Okubo \cite{Okubo} for representing the baryon 20-plet, see
also the Appendix of Ref.~\cite{Liu01}. These
tensors fulfil the conditions

\begin{equation}
\psi_{\mu\nu\lambda} + \psi_{\nu\lambda\mu} + \psi_{\lambda\mu\nu} = 0,
\ \ \psi_{\mu\nu\lambda} = \psi_{\nu\mu\lambda} . 
\end{equation}

In terms of the baryon fields the tensor is given by \cite{Okubo}

\begin{eqnarray}
&& \renewcommand{\arraystretch}{1.3}
\begin{array}{lll} \psi_{111}=p&  \psi_{221}=n&  \psi_{123}= \frac1{\sqrt2} \Sigma^0 \\*
\psi_{231}=\frac1{2} (-\frac1{\sqrt2} \Sigma^0  +\sqrt{\frac{3}{2}}\Lambda) &   
\psi_{312}=\frac1{2} (-\frac1{\sqrt2} \Sigma^0  -\sqrt{\frac{3}{2}}\Lambda) &
\psi_{113}=\Sigma^+ \\*
\psi_{223}=\Sigma^-  & \psi_{331}=\Xi^0&   \psi_{332}=\Xi^- \\
  \nonumber\\
\psi_{124}= \frac1{\sqrt2} \Sigma_c^+ & 
\psi_{241}=\frac1{2} (-\frac1{\sqrt2} \Sigma_c^+  +\sqrt{\frac{3}{2}}\Lambda_c^+) & 
\psi_{412}=\frac1{2} (-\frac1{\sqrt2} \Sigma_c^+  -\sqrt{\frac{3}{2}}\Lambda_c^+) \\*
\psi_{114}=\Sigma_c^{++} & \psi_{224}=\Sigma_c^0 & \psi_{134}=\frac1{\sqrt2}\Xi_c^+ \\*
\psi_{341}=\frac1{2} (-\frac1{\sqrt2} \Xi_c^+  -\sqrt{\frac{3}{2}}\Xi_c'^+) & 
\psi_{413}=\frac1{2} (-\frac1{\sqrt2} \Xi_c^+  +\sqrt{\frac{3}{2}}\Xi_c'^+) &   
\psi_{234}=\frac1{\sqrt2}\Xi_c^0 \\*
\psi_{342}=\frac1{2} (-\frac1{\sqrt2} \Xi_c^0  -\sqrt{\frac{3}{2}}\Xi_c'^0) & 
\psi_{423}=\frac1{2} (-\frac1{\sqrt2} \Xi_c^0  +\sqrt{\frac{3}{2}}\Xi_c'^0) &
\psi_{334}=\Omega_c^0\\
    \nonumber\\
 \psi_{441}=\Xi_{cc}^{++}& \psi_{442}=\Xi_{cc}^+& \psi_{443}=\Omega_{cc}^+ \,.
    \end{array}
\end{eqnarray}
\renewcommand{\arraystretch}{1.0}

The SU(4) 15-plet of the mesons is represented by the tensor

\begin{eqnarray} &&
\begin{array}{llll}
M^1_1=\frac{\pi^0}{\sqrt 2}+\frac{\eta_8}{\sqrt 6}+\frac{\eta_{15}}{\sqrt {12}}
& M^2_1=\pi^+ & M^3_1=K^{+} & M^4_1={\bar D}^{0} \\
M^1_2=\pi^- & M^2_2=-\frac{\pi^0}{\sqrt 2}+\frac{\eta_8}{\sqrt 6}+\frac{\eta_{15}}
{\sqrt {12}} & M^3_2=K^{0} & M^4_2=D^{-} \\
M^1_3=K^{-} & M^2_3={\bar K}^{0} & M^3_3=-\sqrt {\frac{2}{3}}\eta_8+\frac{\eta_{15}}{\sqrt {12}}
& M^4_3=D_s^{-} \\
M^1_4=D^{0} & M^2_4=D^{+} & M^3_4=D_s^{+} & M^4_4=-\frac{3\eta_{15}}{\sqrt {12}}
\end{array}
\;. \nonumber
\end{eqnarray}

Note that the structure for vector mesons is identical and,
therefore, we don't give its form explicitly. It can be obtained via the
obvious replacements $\pi \to \rho$, $K \to K^*$, etc. 

In terms of those tensors the SU(4) invariant interaction Lagrangian is
given formally by 
\begin{equation}
{\mathcal L} = g 
(a \psi^{* \alpha\mu\nu} M^\beta_\alpha \psi_{\beta\mu\nu} + 
 b \psi^{* \alpha\mu\nu} M^\beta_\alpha \psi_{\beta\nu\mu})  
\label{SU4} 
\end{equation}

In the actual evaluation of the baryon-baryon-meson coupling
constants for the SU(4) case we take as reference the standard
SU(3) calculation. There those couplings are obtained from \cite{Polinder}
\begin{eqnarray}
{\mathcal L}&=&\left<\frac{D}{2}\bar{B} \left\{M,B\right\} +\frac{F}{2}\bar{B}\left[M,B\right] \right> \ ,
\label{SU3}
\end{eqnarray}
where $B$ and $M$ are the baryon and meson octets in the usual 
matrix representation \cite{Polinder} and the brackets $\left< ... \right>$ denote
that the trace has to be taken. The two independent coupling constants $F$ and $D$ are
usually expressed by the ratios $\alpha_{ps} = F / (F+D)$ and $1-\alpha_{ps}$, respectively. 
The SU(3) relations for the coupling constants can be read off by re-grouping the 
terms that arise in the explicit evaluation of Eq.~(\ref{SU3}) 
into multiplets within the isospin basis, cf. Eq.~(2.17) in Ref.~\cite{Polinder}. 
The expressions based on the SU(4) Lagrangian (\ref{SU4}) can be mapped onto our SU(3)
results with $a = (-4 + 10\alpha_{ps}) \frac{4}{9}$ and  
$b = (-5 + 8\alpha_{ps}) \frac{4}{9}$. 

The coupling constants at the baryon-baryon-meson vertices relevant for the
present study are given by 
\begin{eqnarray} &&
\renewcommand{\arraystretch}{1.3}
\begin{array}{lll}
g_{\Xi_c\Xi_c \pi} = \alpha_{ps} g_{NN\pi} & 
g_{\Xi_c'\Xi_c' \pi} = \frac{5\alpha_{ps}-2}{3} g_{NN\pi} \\* 
\nonumber
g_{\Xi_c'\Xi_c \pi} = \frac1{\sqrt3}(\alpha_{ps}-1) g_{NN\pi} & \\*
\nonumber
g_{\Xi_c\Xi_c \eta_8} = -\frac1{\sqrt3} \alpha_{ps} g_{NN\pi} & 
g_{\Xi_c'\Xi_c' \eta_8} = \frac1{3\sqrt3} (2- 5\alpha_{ps}) g_{NN\pi} \\*
\nonumber
g_{\Xi_c'\Xi_c \eta_8} = (\alpha_{ps}-1) g_{NN\pi} & \\*
\nonumber
g_{\Xi_c\Xi_c \eta_{15}} = \frac1{\sqrt6} (3-4 \alpha_{ps}) g_{NN\pi} \ & 
g_{\Xi_c'\Xi_c' \eta_{15}} = \frac1{3\sqrt6} (-7+ 4\alpha_{ps}) g_{NN\pi} \\*
\nonumber
g_{\Xi_c\Lambda_c K} = \sqrt{\frac{2}{3}} (\alpha_{ps}-1) g_{NN\pi} & 
g_{\Xi_c'\Lambda_c K} =\frac{\sqrt2}{3} (2- 5\alpha_{ps}) g_{NN\pi} \\*
\nonumber
g_{\Xi_c\Sigma_c K} = \sqrt2 \alpha_{ps} g_{NN\pi} & 
g_{\Xi_c'\Sigma_c K} =\sqrt{\frac{2}{3}} (1- \alpha_{ps}) g_{NN\pi} \\*
\nonumber
& \\*
\nonumber
g_{\Lambda_c\Sigma_c \pi} = \frac{2}{\sqrt3} (1-\alpha_{ps}) g_{NN\pi} & 
g_{\Sigma_c\Sigma_c \pi} = 2 \alpha_{ps} g_{NN\pi} \\*
\nonumber
g_{\Lambda_c\Lambda_c \eta_8} = \frac{2\sqrt3}{9} (-2+5\alpha_{ps}) g_{NN\pi} \ & 
g_{\Lambda_c\Lambda_c \eta_{15}} = \frac1{3\sqrt6} (-7+4\alpha_{ps}) g_{NN\pi} \\*
\nonumber
g_{\Sigma_c\Sigma_c \eta_8} = \frac{2}{\sqrt3} \alpha_{ps} g_{NN\pi} & 
g_{\Sigma_c\Sigma_c \eta_{15}} = \frac1{\sqrt6} (3-4\alpha_{ps}) g_{NN\pi} \\*
\nonumber
g_{\Lambda_c N D} = -\frac1{\sqrt3} (1+2\alpha_{ps}) g_{NN\pi} & 
g_{\Sigma_c N D} = (1-2 \alpha_{ps}) g_{NN\pi}  
\end{array}
\label{couppi} 
\end{eqnarray}
\renewcommand{\arraystretch}{1.0}

In case of pseudoscalar mesons the ratio $\alpha_{ps}$ is fixed from the
non-relativistic quark model (SU(6)), i.e. $\alpha_{ps} = 2/5$ 
\cite{Holzenkamp89}. The contribution of the $\eta$ meson have been
neglected \cite{MHE,Holzenkamp89}. 

For the isoscalar vector mesons $\omega$, $\phi$, and $J/\psi$ we assume 
ideal mixing of the $\omega_{15}$, $\omega_8$ and $\omega_1$ states, i.e.
\begin{eqnarray}
\nonumber
\omega &=& \sqrt{\frac1{2}}\omega_1+\sqrt{\frac1{3}}\omega_8+\sqrt{\frac1{6}}\omega_{15}, \\ 
\nonumber
\phi&=& -\frac1{2}\omega_1+\sqrt{\frac{2}{3}}\omega_8-\sqrt{\frac1{12}}\omega_{15}, \\ 
J/\psi&=& \phantom{-}\frac1{2}\omega_1-\frac{\sqrt3}{2}\omega_{15},  
\end{eqnarray}
and fix the coupling constant of the SU(4) singlet by imposing the OZI rule so 
that $g_{NN\phi}=0$. This also ensures that $g_{NN\,J/\psi}=0$.

For the vector coupling constant the $F/(F+D)$ ratio $\alpha^e_V=1$ is 
used which then yields the following relations for the $\omega$ coupling constants:
\begin{eqnarray}
\nonumber
g_{\Lambda\Lambda\omega}&=&g_{\Sigma\Sigma\omega} = 
g_{\Lambda_c\Lambda_c\omega}=g_{\Sigma_c\Sigma_c\omega} = {2\over 3} g_{NN\omega}, \\
g_{\Xi\Xi\omega}&=&
g_{\Xi_c\Xi_c\omega}=g_{\Xi_c'\Xi_c'\omega} = {1\over 3} g_{NN\omega}, 
\end{eqnarray}

\begin{eqnarray}
\nonumber
g_{\Lambda\Lambda\phi}&=&g_{\Sigma\Sigma\phi} =-{\sqrt{2}\over 3} g_{NN\omega}, \\
\nonumber
g_{\Xi\Xi\phi}&=&
2g_{\Xi_c\Xi_c\phi}=2g_{\Xi_c'\Xi_c'\phi} =-{2\sqrt{2}\over 3} g_{NN\omega}, \\
\nonumber
g_{\Lambda_c\Lambda_cJ/\psi}&=&g_{\Sigma_c\Sigma_cJ/\psi} = {\sqrt{2}\over 3} g_{NN\omega}, \\
g_{\Xi_c\Xi_cJ/\psi}&=&g_{\Xi_c'\Xi_c'J/\psi} = {\sqrt{2}\over 3} g_{NN\omega}, 
\end{eqnarray}

In case of the tensor coupling constants $f$ the SU(3) relations are actually applied
to the combination of the electric and magnetic coupling, $G= g + f$, and with 
the $F/(F+D)$ ratio $\alpha^m_V=2/5$ \cite{Holzenkamp89}. Taking also into account
that in the full Bonn $NN$ model one has $f_{NN\omega}$=0 \cite{MHE}
yields then the following relations for the $f$'s: 

\begin{eqnarray}
\nonumber
f_{\Lambda\Lambda\omega}&=&f_{\Lambda_c\Lambda_c\omega} = -{1\over 2} f_{NN\rho}, \\
\nonumber
f_{\Sigma\Sigma\omega}&=&f_{\Sigma_c\Sigma_c\omega} = +{1\over 2} f_{NN\rho}, \\
f_{\Xi\Xi\omega}&=&-2f_{\Xi_c\Xi_c\omega} = 2f_{\Xi_c'\Xi_c'\omega} = -{1\over 2} f_{NN\rho}
\end{eqnarray}

\begin{eqnarray}
\nonumber
f_{\Lambda\Lambda\phi}&=&-f_{\Sigma\Sigma\phi} = -{\sqrt{2}\over 2} f_{NN\rho}, \\
\nonumber
f_{\Xi\Xi\phi}&=&2f_{\Xi_c\Xi_c\phi} =-2f_{\Xi_c'\Xi_c'\phi} = -{\sqrt{2}\over 2} f_{NN\rho} \\
\nonumber
f_{\Lambda_c\Lambda_c J/\psi}&=&-f_{\Sigma_c\Sigma_c J/\psi} = {\sqrt{2}\over 2} f_{NN\rho}, \\
f_{\Xi_c\Xi_cJ/\psi} &=&-f_{\Xi_c'\Xi_c'J/\psi} = -{\sqrt{2}\over 2} f_{NN\rho}
\end{eqnarray}

In the study of strangeness production 
\cite{Haidenbauer:1991,Haidenbauer:1992,Haidenbauer:1993,Haidenbauer:XX} 
the contribution of $\phi$ meson exchange was ignored. Since its contribution is
of rather short range it was argued that it is effectively included via the real 
part of the phenomenological annihilation potential, which is also of short
range and has to determined by a fit to data anyway. We adopt the same point of
view here, and we also omit the contribution of the even shorter ranged
contribution from $J/\psi$ exchange. Exploratory calculations for 
strangeness production with inclusion of $\phi$ exchange resulted in an
increase of the cross sections by a factor of roughly 2. However, as expected
this increase can be easily compensated by an appropriate adjustment of the 
parameters in the annihilation potential so that one arrives again
at results that agree with the measurements.
A corresponding compensation takes place also in the charm sector if we include
the $\phi$ meson but then adopt likewise the re-adjusted parameters (from the
strangeness sector) for the final-state interaction in $\Lambda_c\bar\Lambda_c$, etc. 

In the works of the Bonn-J\"ulich groups the $\sigma$ meson stands for the correlated
$\pi\pi$ s-wave interaction and is neither considered to be an SU(3) singlet nor
a member of the $0^+$-meson octet. Here, for simplicity reasons we simply take
over the coupling constants used at the $\Lambda\Lambda\sigma$-, 
$\Sigma\Sigma\sigma$-, and $\Xi\Xi \sigma$ vertices in previous works 
\cite{Holzenkamp89,Haidenbauer:XX}
for the corresponding vertices for charmed baryons. 
Table~\ref{param} summarizes the values of the coupling constants and cutoff masses
of the vertex form factors employed in the present calculation.  

\begin{table}[htb]
\renewcommand{\arraystretch}{1.2}
\centering
\caption{\label{param} 
Coupling constants and cutoff masses at the $\Xi\Lambda K$, $\Xi\Xi\pi$, etc. 
and the corresponding $\Xi_c\Lambda_c K$, $\Xi_c\Xi_c\pi$, etc. vertices. 
The coupling constants are obtained from SU(4) relations with 
$g_{NN\pi}/\sqrt{4\pi}$ = 3.795,
$g_{NN\rho}/\sqrt{4\pi}$ = 0.917, and
$f_{NN\rho}/\sqrt{4\pi}$ = 5.591 and
the $F/(F+D)$ ratios $\alpha_{ps} = 2/5$, 
$\alpha^e_{v} = 1$ and $\alpha^m_{v} = 2/5$. 
}
\vskip 0.2cm 
\begin{ruledtabular}\begin{tabular}{ l ccc cc}
%\hline
%\hline
        & \multicolumn{2}{c}{Strangeness} & & \multicolumn{2}{c}{Charm}\\
Vertex  & $g_{\alpha}/\sqrt{4\pi}$ & $f_{\alpha}/\sqrt{4\pi}$ & $\Lambda_{\alpha}$ (GeV)
& $g_{\alpha}/\sqrt{4\pi}$ & $f_{\alpha}/\sqrt{4\pi}$ \\
\hline
$\Xi\Lambda K  $ & 1.315 &        & 2.0 &-1.859 &       \\
$\Xi\Sigma  K  $ &-3.795 &        & 2.0 & 2.147 &       \\
$\Xi\Lambda K^*$ & 1.588 & 0.666  & 2.2 &       &-3.188 \\
$\Xi\Sigma  K^*$ &-0.917 & -5.591 & 2.2 & 1.297 & 2.385 \\
$\Xi\Xi\pi     $ &-0.759 &        & 1.3 & 1.518 &       \\
$\Xi\Xi\rho    $ & 0.971 & -2.219 & 1.3 & 0.917 & 1.686 \\
$\Xi\Xi\omega  $ & 1.491 & -2.800 & 2.0 & 1.491 & 1.398 \\
$\Xi\Xi\phi    $ &-4.216 & -3.953 &     &-2.108 &-1.977 \\
$\Xi\Xi J/\psi $ &       &        &     & 2.108 &-3.953 \\
$\Xi\Xi\sigma  $ & 3.162 &        & 1.7 & 3.162 &       \\
\hline
$\Xi_c'\Sigma K  $    &  &  & 2.0 & 1.859 &       \\
$\Xi_c'\Lambda K^*$   &  &  & 2.2 & 1.297 &-1.297 \\
$\Xi_c'\Sigma  K^*$   &  &  & 2.2 & 1.682 & 1.682 \\
$\Xi_c'\Xi_c'\rho   $ &  &  & 1.3 & 0.917 &-0.917 \\
$\Xi_c'\Xi_c'\omega $ &  &  & 2.0 & 1.491 &-1.398 \\
$\Xi_c'\Xi_c'\phi   $ &  &  &     &-2.108 & 1.977 \\
$\Xi_c'\Xi_c'J/\psi $ &  &  &     & 2.108 & 3.953 \\
$\Xi_c'\Xi_c'\sigma $ &  &  & 1.7 & 3.162 &       \\
%\hline
\end{tabular}
\end{ruledtabular}\end{table}
\vspace{1.0cm}

%
%%%%%%%%%%%%%%%%%%%%%%%%%%%%%%%%%%%%%%%%%%%%%%%%%%%%%%

\end{document}